\documentclass{jfm}

\usepackage{graphicx, comment}
\usepackage{newtxtext}
\usepackage{newtxmath}
\usepackage{amssymb}
\usepackage[utf8]{inputenc}
\usepackage{natbib}
\usepackage[normalem]{ulem}
\usepackage{bm}
\usepackage{cancel}
\usepackage{hyperref}
\hypersetup{
    colorlinks = true,
    urlcolor   = blue,
    citecolor  = black,
}
\newtheorem{lemma}{Lemma}
\newtheorem{corollary}{Corollary}
\newcommand{\RomanNumeralCaps}[1]

\title{Universal Translational and Rotational Mobility Expressions of Phoretic and Self-phoretic Particles with Arbitrary Interaction Potentials}

\author{Arkava Ganguly$^\mathrm{*}$\aff{1},
  Souradeep Roychowdhury$^\mathrm{*}$\aff{1}
 \and Ankur Gupta\aff{1}\corresp{\email{ankur.gupta@colorado.edu}}}

\affiliation{\aff{1}Department of Chemical and Biological Engineering, University of Colorado Boulder, Boulder, CO 80303, USA}

\begin{document}
\maketitle
\def\thefootnote{*}\footnotetext{These authors contributed equally to this work}
\begin{abstract}
The mobility of externally-driven phoretic propulsion of particles is evaluated by simultaneously solving the solute conservation equation, interaction potential equation, and the modified Stokes equation. While accurate, this approach is cumbersome, especially when the interaction potential decays slowly compared to the particle size. In contrast to external phoresis, the motion of self-phoretic particles is typically estimated by relating the translation and rotation velocities with the local slip velocity. While this approach is convenient and thus widely used, it is only valid when the interaction decay length is significantly smaller than the particle size. Here, by taking inspiration from Brady J. Fluid Mech. (2021), vol. 922, A10, which combines the benefits of two approaches, we reproduce their unified mobility expressions with arbitrary interaction potentials and show that these expressions can conveniently recover the well-known mobility relationships of external electrophoresis and diffusiophoresis for arbitrary double-layer thickness. Additionally, we show that for a spherical microswimmer, the derived expressions relax to the slip velocity calculations in the limit of the thin interaction lengthscales. We also employ the derived mobility expressions to calculate the velocities of an autophoretic Janus particle. We find that there is significant dampening in the translation velocity even when the interaction length is an order of magnitude larger than the particle size. Finally, we study the motion of a catalytically self-propelled particle, while it also propels due to external concentration gradients, and demonstrate how the two propulsion modes compete with each other.

\end{abstract}



\section{Introduction}
\label{sec:introduction}
Phoretic phenomena, i.e., the movement of particles in response to an external field \citep{anderson1989colloid, velegol2016origins, khair2022nonlinear}, is pivotal for a range of applications such as separation of biomacromolecules \citep{heller2001principles,lee2012agarose}, purification of nucleic acids from whole blood \citep{persat2009purification}, measurement of zeta potential \citep{doane2012nanoparticle,shin2017low}, banding of colloidal particles \citep{abecassis2008boosting, banerjee2019long, raj2023two}, membrane-less water filtration \citep{shin2017membraneless} and understanding biological pattern formation \citep{alessio2023diffusiophoresis}, among others. In contrast, self-phoretic particles, also known as microswimmers, respond to a self-generated field gradient \citep{paxton2004catalytic,howse2007self,ebbens2010pursuit,popescu2016self, ganguly2023diffusiophoresis}. Microswimmers are extensively studied for applications in targeted drug delivery \citep{xuan2014self, luo2018micro}, environmental remediation \citep{gao2013seawater,wang2019photocatalytic}, remote sensing of toxic chemicals \citep{esteban2016aptamer}, the autonomous motion of microbots \citep{zarei2018self,hu2022self}, and collective behavior of active colloids \citep{palacci2013living,takatori2016forces,illien2017fuelled}. The mobility expressions of phoretic and self-phoretic processes are identical, with the key distinction being that the origin of field gradients in the two processes is different. In phoretic processes, this field gradient is externally imposed on colloidal particles, while in self-phoretic particles they are locally generated by the particles themselves typically through surface reactions or other mechanisms.
 
Studies on external electrophoretic motion have focused on the dependence of electrophoretic mobility on the effect of particle shape \citep{yoon1989electrophoresis,solomentsev1994electrophoresis}, surface heterogeneity \citep{fair1992electrophoresis,velegol1996probing}, finite double-layer thickness \citep{henry1931cataphoresis, o1978electrophoretic}, and more recently, strong deformation of double-layers \citep{khair2018strong,khair2022nonlinear} and charge reversal \citep{kubivckova2012overcharging, gupta2020ionic}. Similarly, researchers have predicted the dependence of diffusiophoretic mobility \citep{anderson1989colloid,brady2011particle} on finite double-layer thickness \citep{prieve1984motion,keh2000diffusiophoretic}, surface chemistry \citep{gupta2020diffusiophoresis}, and multiple electrolytes \citep{gupta2019diffusiophoretic, alessio2021diffusiophoresis}. Studies on self-phoretic systems \citep{ramaswamy2010mechanics,moran2017phoretic} focus on the impact of particle shape \citep{shklyaev2014non,nourhani2016geometrical,poehnl2020axisymmetric,daddi2021optimal,ganguly2023going, raj2023motion, lee2023magnetically}, active patch shape \citep{lisicki2018autophoretic,lee2021fabrication}, surface interaction \citep{sharifi2013diffusiophoretic} and finite P\'{e}clet number \citep{michelin2014phoretic}. 

Broadly speaking, there are two approaches for predicting the mobilities described above. The first approach solves the coupled solute conservation equations and the modified Stokes equation \citep{henry1931cataphoresis,o1978electrophoretic,prieve1984motion,prieve1987diffusiophoresis,anderson1989colloid,keh2000diffusiophoretic,sharifi2013diffusiophoretic,khair2018strong,khair2022nonlinear, gupta2019diffusiophoretic} and employs a force-free and torque-free condition to arrive at the translation velocity, $\mathbf{U}$, and rotational velocity, $\mathbf{\Omega}$, of the particle. Thus the above approach requires resolving the interaction potential simultaneously with the hydrodynamic equations. While exact and powerful, the methodology described above is cumbersome for analytical results when the particle-fluid interaction potential decays at much larger length scales than the particle size. Further, the solution strategy needs to be revised whenever the interaction potential changes, making it less convenient to be integrated into other analyses.

The second approach employs the reciprocal theorem in the thin interaction length limit \citep{stone1996propulsion,brady2011particle,michelin2014phoretic,lisicki2018autophoretic,masoud2019reciprocal,poehnl2020axisymmetric,ganguly2023going, raj2023motion}. In this limit, it is assumed that there exists a slip velocity, $\mathbf{u}_s$, at the interface of the inner region where the interaction potential is non-zero, and the outer region where the interaction potential is zero. This allows one to treat the outer problem as a classical Stokes flow problem with a slip boundary condition. Consequently, $\mathbf{U}$ and $\mathbf{\Omega}$ can be represented as surface integrals of appropriate functions of $\mathbf{u}_s$. This approach based on the reciprocal theorem was first utilized by \cite{stone1996propulsion} to study the impact of distortions in spherical microswimmers. This methodology is particularly powerful because, unlike the first method, computing $\mathbf{U}$ and $\mathbf{\Omega}$ is relatively straightforward and agnostic to the mechanistic origin of $\mathbf{u}_s$. However, this approach is valid only when the interaction potential length is significantly smaller than the particle size, restricting its applicability. Additionally, it requires the knowledge of $\mathbf{u}_s$ {\it a priori}, and most studies have to rely on a lumped mobility parameter to estimate the value of $\mathbf{u}_s$ \citep{michelin2014phoretic, lisicki2018autophoretic, poehnl2021phoretic, ganguly2023going, raj2023motion}.

In this work, we seek to unify the benefits realized through the two approaches by employing the results of ~\cite{brady2021phoretic} and demonstrating how it can reconcile a large volume of mobility results for externally-driven and self-phoretic propulsion of particles, and using these results for additional analyses. This approach is similar to the prior literature on the inertial correction to Stokes flow \citep{ brenner1963resistance,hinch1991, leal2007advanced}, and swimming through non-Newtonian fluids \citep{datt2015squirming, elfring2016effect, datt2017active}, where the first-order corrections include a body force term from the leading order and reciprocal theorem is employed to find the resulting motion. In section \ref{sec:deriv}, we obtain a general mobility expression for an arbitrary particle shape, subjected to an osmophoretic (combination of osmotic and phoretic force) body force $\mathbf{b}$, identical to the results in \cite{brady2021phoretic}. Subsequently, we take our expression to the thin interaction length scale limit and retrieve the mobility expressions in \cite{stone1996propulsion}, see section \ref{sec:valid}. In section \ref{sec:valid}, we also retrieve the expression for both the electrophoretic mobility of translation of a charged spherical particle in an external electric field, obtained by \cite{henry1931cataphoresis}.  Additionally, we derive the diffusiophoretic mobility of a charged spherical particle in an externally imposed solute gradient at finite interaction lengths, as first obtained by \cite{keh2000diffusiophoretic}. Finally in section \ref{sec: autophoretic motion of microswimmers}, we apply our result to study the autophoretic motion of spherical microparticles with catalytic caps. We study how translation velocity depends on the cap size, the surface interaction potential, and the interaction length relative to particle size. Since our methodology works for both externally driven and self-propelling particles, we also study particle propulsion by both modes simultaneously; see section~\ref{sec:swimming}.  These model problems demonstrate the wide applicability of the expressions derived by \cite{brady2021phoretic} and reproduced in this manuscript. Finally, in section \ref{sec: conclusion} we summarize the key findings of our work and outline future ideas.
 
\section{Derivation of the unified mobility expression}
\label{sec:deriv}
\begin{figure}
    \centering
    \includegraphics[width=0.8\textwidth]{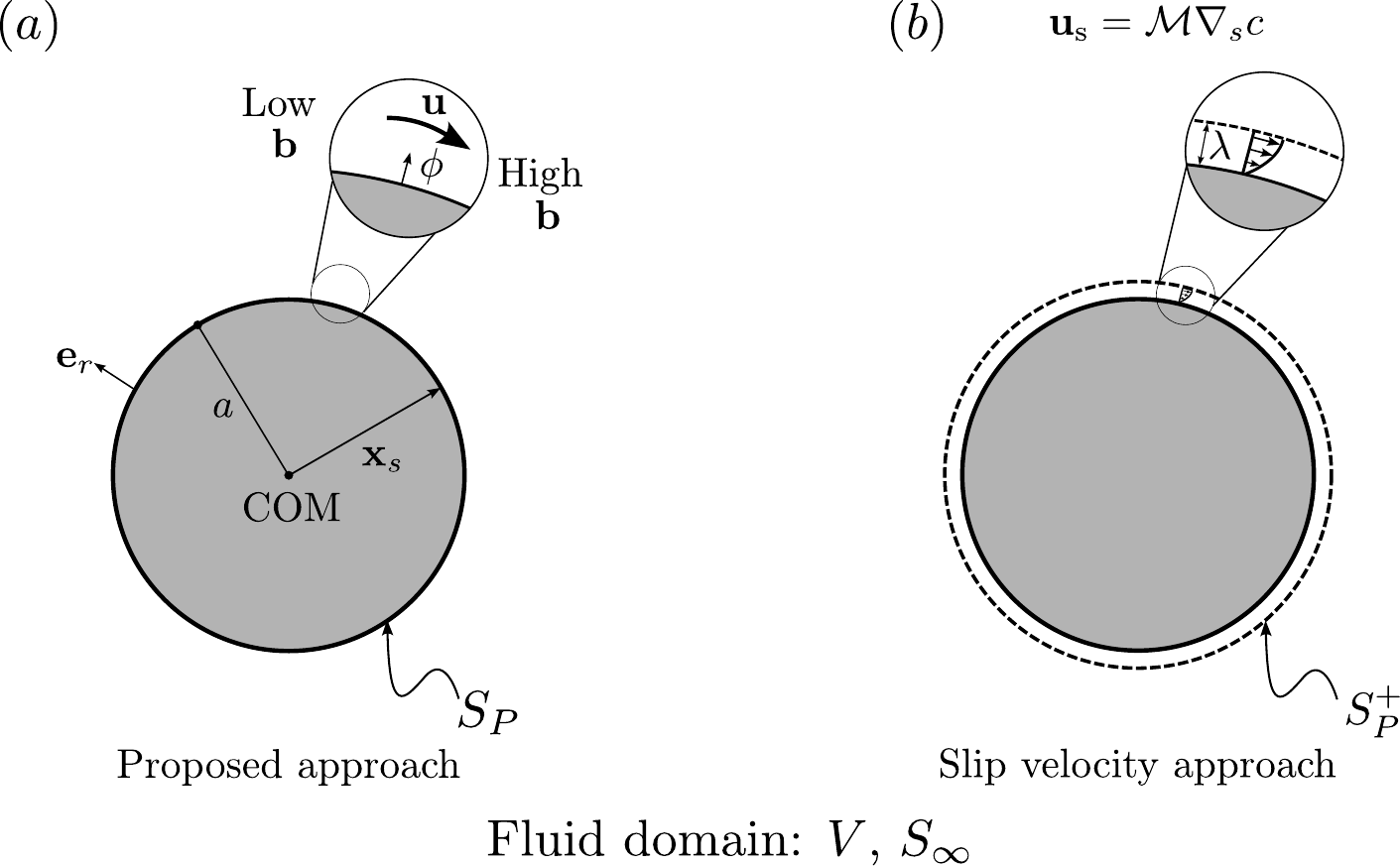}
    \caption{Two approaches to finding the velocity of a particle by resolving the fluid velocity at the particle surface. (a) Obtain the fluid velocity near the particle surface by resolving the modified Stokes equation with an arbitrary body force ($\mathbf{b}$). The body force, $\mathbf{b}$, depends on charge ($\rho$), salt ($s$), and interaction potential ($\phi$). (b) When the interaction length is small $\lambda/a \ll 1$, the velocity near the fluid surface, at the outer edge of the interaction layer, $\mathbf{u}_\mathrm{s}$, is taken to be the velocity at the particle surface. The slip velocity, $\mathbf{u}_\mathrm{s}$, depends on the lumped mobility ($\mathcal{M}$), which depends on the interaction between the surface and solute, and the solute concentration at the vicinity of the surface ($c$).}
    \label{fig:thin_slip_finite_interaction_length__comparison}
\end{figure}

In this section, we derive the translation velocity ($\mathbf{U}$) and rotational velocity ($\mathbf{\Omega}$) of an arbitrary particle with surface $S_p$ immersed in a fluid of volume $V$ due to an arbitrary osmophoretic body force $\mathbf{b}$, see Fig \ref{fig:thin_slip_finite_interaction_length__comparison}a. The particle surface is defined via the vector $\mathbf{x}_S$ relative to the center of mass of the particle. We define $\mathbf{e}_r$ as the outward unit normal to the particle surface, $r$ is the distance from the center of mass of the particle and $\mathbf{r}$ is the position vector defined from the center of mass of the particle. The fluid velocity around the particle can be resolved through the modified Stokes equation, defined as,
\begin{equation}
    \bm{\nabla}\cdot\bm{\sigma} + \mathbf{b} = \bm{0}, \label{Eq: modified stokes}
\end{equation}
where $\bm{\sigma}$ is the hydrodynamic stress tensor. The velocity field $\mathbf{u}$ is assumed to decay to zero in the far-field, $\mathbf{u}|_{r \to \infty} \to 0$. At the particle surface, the fluid obeys a no-slip, rigid body boundary condition, $\mathbf{u}|_{S_p} = \mathbf{U} + \mathbf{\Omega} \times \mathbf{x}_{S}$.
To obtain $\mathbf{U}$ and $\mathbf{\Omega}$ using Lorentz reciprocal theorem \citep{masoud2019reciprocal}, we define an auxiliary Stokes flow ($\hat{\mathbf{U}},\hat{\mathbf{\Omega}}$) while preserving particle geometry with the same no-slip rigid surface, $\hat{\mathbf{u}}|_{S_p} = \hat{\mathbf{U}} + \hat{\mathbf{\Omega}} \times \mathbf{x}_{S}$, and far-field decay, $\hat{\mathbf{u}}|_{r \to\infty }\to 0$, boundary conditions. 

Using the Lorentz reciprocal theorem, we can relate the phoretic problem ($\mathbf{U}$,$\mathbf{\Omega}$,$\mathbf{b}$) and the auxiliary problem ($\hat{\mathbf{U}}$,$\hat{\mathbf{\Omega}}$,$\hat{\mathbf{b}}$) to be
\begin{equation}
    \displaystyle \int_{S_p} \mathbf{e}_r\cdot\bm{\sigma}\cdot\hat{\mathbf{u}}\;{\rm dS} - \displaystyle \int_{S_p} \mathbf{e}_r\cdot\hat{\bm{\sigma}}\cdot\mathbf{u}\;{\rm dS} =\displaystyle \int_V \hat{\mathbf{u}}\cdot\mathbf{b}\;{\rm dV} - \displaystyle \int_V \mathbf{u}\cdot\hat{\mathbf{b}}\;{\rm dV}. \label{Eq: lorentz-reciprocal-thm}
\end{equation}
Substituting in the expressions of the fluid velocities, $\mathbf{u}|_{S_p}$ and $\hat{\mathbf{u}}|_{S_p}$, at the particle surface we can simplify  Eq. \eqref{Eq: lorentz-reciprocal-thm}. Additionally, we assume that there is no body force in the auxiliary problem, ${\mathbf{\hat b}} = \bm{0}$. Thus we can rewrite Eq. \eqref{Eq: lorentz-reciprocal-thm} to be,
\begin{equation}
    \displaystyle \int_{S_p} \mathbf{e}_r\cdot\bm{\sigma}\cdot\left(\hat{\mathbf{U}}+\hat{\mathbf{\Omega}}\times \mathbf{x}_S\right) {\rm d S} - \displaystyle \int_{S_p} \mathbf{e}_r\cdot\hat{\bm{\sigma}}\cdot\left(\mathbf{U}+\mathbf{\Omega}\times\mathbf{x}_S\right) {\rm d}S = \displaystyle \int_V \hat{\mathbf{u}}\cdot\mathbf{b}\; {\rm dV}. \label{Eq: lorentz-reciprocal-thm-2}
\end{equation}
Since the inertia of the particle is negligible, for both the phoresis and the auxiliary problem, the particle is force and torque-free.  For the phoretic propulsion, the hydrodynamic force and torque is balanced by the osmophoretic force and torque, or
\begin{equation}
    \underbrace{\displaystyle \int_{S_p} \mathbf{e}_r\cdot\bm{\sigma}\; {\rm dS}}_\text{hydrodynamic} \  \ \  \ \underbrace{-  \displaystyle \int_V \mathbf{b}\; {\rm dV}}_\text{osmophoretic} = \mathbf{0}, \label{Eq: phoretic_trans_stress_constraint}
\end{equation}
\begin{equation}
    \underbrace{\displaystyle \int_{S_p} \mathbf{x}_s \times \mathbf{e}_r \cdot \bm{\sigma} \;{\rm dS}}_\text{hydrodynamic}  \ \  \ \  \underbrace{- \displaystyle \int_V \mathbf{r}\times\mathbf{b}\; {\rm dV}}_\text{osmophoretic} = \mathbf{0}, \label{Eq: phoretic_rot_stress_constraint}
\end{equation}
where the negative sign in front of the osmophoretic term comes because the osmophoretic force on the particle is equal in magnitude to the osmophoretic force on the fluid but opposite in sign \citep{brady2011particle}. For the auxiliary system, we balance the hydrodynamic and external forces and torques, or

\begin{equation}
    \underbrace{\displaystyle \int_{S_p} \hat{\bm{\sigma}}\cdot \mathbf{e}_r\;{\rm dS} }_\text{hydrodynamic} \ \  \ \ + \underbrace{\hat{\mathbf{F}}_{\rm ext}}_\text{external}= \bm{0}, \label{Eq: force-free aux prob}
\end{equation}
\begin{equation}
    \underbrace{\displaystyle \int_{S_p} \mathbf{x}_s \times \hat{\bm{\sigma}}\cdot\mathbf{e}_r\;{\rm dS }}_\text{hydrodynamic} \ \  \ \  + \underbrace{\hat{\mathbf{L}}_{\rm ext}}_\text{external}= \bm{0}, \label{Eq: torque-free aux prob}
\end{equation}
where $\hat{\mathbf{F}}_{\rm ext}$ and $\hat{\mathbf{L}}_{\rm ext}$ are the external force and torque required to move the particle in the auxiliary problem, which is to be determined. We note that Eqs. \eqref{Eq: phoretic_trans_stress_constraint}-\eqref{Eq: phoretic_rot_stress_constraint} are different from Eqs. \eqref{Eq: force-free aux prob}-\eqref{Eq: torque-free aux prob} since the phoretic propulsion is induced by $\mathbf{b}$ whereas in the auxiliary problem motion is caused by $\hat{\mathbf{F}}_{\rm ext}$ and $\hat{\mathbf{L}}_{\rm ext}$. 

To calculate $\mathbf{U}$ and $\mathbf{\Omega}$ for a given $\mathbf{b}$ from Eq. \eqref{Eq: lorentz-reciprocal-thm-2} - \eqref{Eq: torque-free aux prob}, we need to express, $\hat{\mathbf{F}}_{\rm ext}$ , $\hat{\mathbf{L}}_{\rm ext}$ and $\mathbf{\hat{u}}$ in the auxiliary problem as functions of $\mathbf{\hat{U}}$ and $\mathbf{\hat{\Omega}}$. To do so, we use a resistance formulation  to write,

\begin{equation}
    \begin{bmatrix}
        \hat{\mathbf{F}}_{\rm ext}\\
        \hat{\mathbf{L}}_{\rm ext}
    \end{bmatrix} = \begin{bmatrix}
        \mathbf{R}_{FU} & \mathbf{R}_{F\Omega} \\
        \mathbf{R}_{LU} & \mathbf{R}_{L\Omega}
    \end{bmatrix}\cdot\begin{bmatrix}
        \hat{\mathbf{U}} \\
        \hat{\mathbf{\Omega}}
    \end{bmatrix}, \label{Eq: resistance_formulation}
\end{equation}
where the resistance matrices $\mathbf{R}_{FU}$, $\mathbf{R}_{F\Omega}$, $\mathbf{R}_{LU}$, and $\mathbf{R}_{L\Omega}$ relate the driving force ($\hat{\mathbf{F}}_{\rm ext}$) and torque ($\hat{\mathbf{L}}_{\rm ext}$) to the translational  ($\mathbf{\hat{U}}$) and rotational velocity ($\mathbf{\hat{\Omega}}$). Further, we describe $\hat{\mathbf{u}}$ as 
\begin{equation}
\hat{\mathbf{u}}=\mathbf{D}_T\cdot\hat{\mathbf{U}}+\mathbf{D}_R\cdot\hat{\mathbf{\Omega}}\times\mathbf{r}, \label{Eq: u_hat_disturbance}
\end{equation}
where $\mathbf{D}_T$ is the translation disturbance tensor and $\mathbf{D}_R$ is the rotation disturbance tensor. Eqs. \eqref{Eq: resistance_formulation} - \eqref{Eq: u_hat_disturbance}  combined with Eqs. \eqref{Eq: force-free aux prob} - \eqref{Eq: torque-free aux prob} provide necessary information to simplify Eq. \eqref{Eq: lorentz-reciprocal-thm-2} as a function of $\mathbf{\hat{U}}$ and $\mathbf{\hat{\Omega}}$. 

Next, we choose convenient values of $\mathbf{\hat{U}}$ and $\mathbf{\hat{\Omega}}$ to simplify Eq. \eqref{Eq: lorentz-reciprocal-thm-2}. Specifically, we use six auxiliary flow problems, pure translation ($\hat{\mathbf{\Omega}} = \mathbf{0}$ and $\hat{\mathbf{U}}$ = $U_0 \mathbf{e}_1$, $U_0 \mathbf{e}_2$, $U_0 \mathbf{e}_3$) and pure rotation ($\hat{\mathbf{U}}=\mathbf{0}$ and $\hat{\mathbf{\Omega}}=U_0/a \mathbf{e}_1$, $U_0/a \mathbf{e}_2$, $U_0/a \mathbf{e}_3$) with $U_0$ being the characteristic velocity scale and $a$ being the characteristic particle length, to obtain
\begin{equation}
    \begin{bmatrix}
        \mathbf{R}_{FU} & \mathbf{R}_{F\Omega} \\
        \mathbf{R}_{LU} & \mathbf{R}_{L\Omega}
    \end{bmatrix}\cdot\begin{bmatrix}
        {\mathbf{U}} \\
        {\mathbf{\Omega}}
    \end{bmatrix} = \begin{bmatrix}
        \displaystyle \int_V \left(\mathbf{D}_T - \mathbf{I}\right)\cdot\mathbf{b}\;{\rm dV} \\
        \displaystyle \int_V \left(\mathbf{D}_R - \mathbf{I}\right)\cdot\mathbf{r}\times\mathbf{b}\;{\rm dV}
    \end{bmatrix} , \label{Eq: resistance_volume_integral}
\end{equation}

By inverting the resistance tensor, we obtain a mobility formulation that resolves $\mathbf{U}$ and $\mathbf{\Omega}$ in terms of volume integrals of $\mathbf{b}$,

\begin{equation}
    \begin{bmatrix}
        \mathbf{U} \\
        \mathbf{\Omega}
    \end{bmatrix} = \begin{bmatrix}
        \mathbf{M}_{UF} & \mathbf{M}_{UL} \\
        \mathbf{M}_{\Omega F} & \mathbf{M}_{\Omega L}
    \end{bmatrix}\cdot\begin{bmatrix}
        \displaystyle \int_V \left(\mathbf{D}_T - \mathbf{I}\right)\cdot\mathbf{b}\;{\rm dV} \\
        \displaystyle \int_V \left(\mathbf{D}_R - \mathbf{I}\right)\cdot\mathbf{r}\times\mathbf{b}\;{\rm dV}
    \end{bmatrix} \label{Eq: unified_mobility_expression}
\end{equation}
where the matrices $\mathbf{M}_{FU}$, $\mathbf{M}_{F\Omega}$, $\mathbf{M}_{LU}$, and $\mathbf{M}_{L\Omega}$ are the corresponding mobility tensors. For an in-depth mathematical analysis and mechanistic discussion regarding the various forms of $\mathbf{b}$, we redirect the reader to \cite{brady2021phoretic}.

Physically, Eq. \eqref{Eq: unified_mobility_expression} is insightful as it helps parse apart the difference between the phoretic problem and the auxiliary problem. The rightmost term is the effective force and torque on the particle due to phoretic interactions and has two contributions: (1) the term associated with the identity tensor ($\mathbf{I}$) is the osmophoretic force and torque acting on the particle, and (2) the term associated with the disturbance tensors ($\mathbf{D}_T$, $\mathbf{D}_R$) is the hydrodynamic correction to the distribution of body forces around the particle.  This correction arises because the phoretic interactions near the particle surface lead to an additional compensating fluid motion~\citep{brady2011particle} causing a long-range hydrodynamic disturbance. This effect is not captured in the definition of the hydrodynamic mobility tensor and thus manifests separately. If the terms associated with disturbance tensors were not present, Eq. \eqref{Eq: unified_mobility_expression} is essentially identical to Eq. \eqref{Eq: resistance_formulation} with osmophoretic force on the particle as the external force.%

We note that Eq. \eqref{Eq: unified_mobility_expression} takes an explicit form only when $\mathbf{b}$ is independent of $\mathbf{U}$ and $\mathbf{\Omega}$ and is thus most convenient for systems with small P{\'e}clet number ($\mathrm{Pe} \ll 1$), where $\mathrm{Pe} = {U_0 L}/{D}$, and $D$ is the diffusivity of the solute. The distinction from prior work, such as \cite{stone1996propulsion,michelin2014phoretic,lisicki2018autophoretic,poehnl2020axisymmetric,poehnl2021phoretic,ganguly2023going}, that utilize Lorentz reciprocal theorem is that they invoke the thin interaction length limit and apply the analysis in the outer region where $\mathbf{b}=\bm{0}$; see Fig. \ref{fig:thin_slip_finite_interaction_length__comparison}b. Consequently, they do not arrive at Eq. \eqref{Eq: unified_mobility_expression} but rather represent $\mathbf{U}$ and $\mathbf{\Omega}$ in terms of a slip velocity at the particle surface $\mathbf{u}_s$.

We acknowledge that similar results have been presented in \cite{khair2018strong} and \cite{brady2021phoretic}. However, in \cite{khair2018strong}, $\mathbf{b}$ only focused on the electrophoretic contributions. In contrast, \cite{brady2021phoretic} argued that $\mathbf{b}$ should include both osmotic and phoretic contributions, and we thus refer to $\mathbf{b}$ as an osmophoretic body force.  Care should be taken that the osmotic contribution only includes excess osmotic  effect since a particle cannot move without a phoretic interaction; interested readers are referred to ~\cite{brady2021phoretic}. The phoretic contribution arises from the interaction of the particle with a macroscopically established potential field. The nature of this field depends on the specific model problem under consideration. We refer the readers to \cite{brady2021phoretic} for an in-depth mathematical analysis and a general discussion on the mechanistic origin of $\mathbf{b}$. Building on the work by \cite{brady2021phoretic}, we systematically illustrate how both phoretic and osmotic contributions to the body force term are required to reconcile a broad range of results in the literature and arrive at universal mobility relationships. Additionally, through this framework, we quantify the impact of interaction length on microswimmer motion in electrolytic solutions, elaborating on the suggestion established in \cite{brady2021phoretic}. 

For a spherical particle (see \cite{duprat2016fluid} for derivation), the relevant hydrodynamic parameters are $\mathbf{D}_T = \frac{3a}{4r}\left(\mathbf{I}+\mathbf{e}_r\mathbf{e}_r\right)+\frac{a^3}{4r^3}\left(\mathbf{I}-3\mathbf{e}_r\mathbf{e}_r\right)$, $\mathbf{D}_R = \frac{a^3}{r^3}\mathbf{I}$, $\mathbf{M}_{UF} = \frac{1}{6\pi\mu a}\mathbf{I}$, $\mathbf{M}_{UL}=0$, $\mathbf{M}_{\Omega F}=0$ and $\mathbf{M}_{\Omega L} = \frac{1}{8\pi\mu a^3} \mathbf{I}$, where $a$ is the radius of the sphere, $\mu$ is the fluid viscosity, $r$ is the radial distance from the center of the sphere, and $\mathbf{e}_r$ is the radial vector pointing away from the center. Substituting, these definitions of the hydrodynamic disturbance and mobility in Eq. \eqref{Eq: unified_mobility_expression} we obtain,
\begin{equation}
    \mathbf{U} = \frac{1}{6\pi\mu a} \displaystyle \int_V \left[\left(\frac{3a}{2r}-\frac{a^3}{2r^3}-1\right)\mathbf{b}_\perp + \left(\frac{3a}{4r}+\frac{a^3}{4r^3}-1\right)\mathbf{b}_\parallel\right]\;{\rm dV}, \label{Eq: unified_trans_exp_for_spheres}
\end{equation}
\begin{equation}
    \mathbf{\Omega} = \frac{1}{8\pi\mu a^3} \displaystyle \int_V r\left(\frac{a^3}{r^3}-1\right)\mathbf{e}_r \times \mathbf{b}_\parallel\;{\rm dV}, \label{Eq: unified_rot_exp_for_spheres}
\end{equation}
where the body force is decomposed into $\mathbf{b}=\mathbf{b}_\perp+\mathbf{b}_\parallel$. The perpendicular subscript denotes the component normal to the sphere and the parallel subscript denotes the component parallel to the surface. Eq. \eqref{Eq: unified_trans_exp_for_spheres}
 - \eqref{Eq: unified_rot_exp_for_spheres} was also reported in prior literature for phoretic systems ~\citep{brady2021phoretic} as well as for different physical systems \citep{brenner1963resistance, hinch1991, leal2007advanced, datt2015squirming,elfring2016effect,datt2017active}. We extensively validate this result in the next section and show it relaxes to the various well-known expressions present in the literature, for both microswimmers and externally-driven particles. Further, we employ this expression to study a microswimmer in the arbitrary interaction layer limit and a microswimmer driven by an external gradient in addition to its self-propelling mode of swimming.
 
 \section{Validation}
\label{sec:valid}
\subsection{Simplification at the limit of the thin interaction length scale}

In this subsection, we aim to simplify Eq. \eqref{Eq: unified_trans_exp_for_spheres} - \eqref{Eq: unified_rot_exp_for_spheres} in the limit of the thin interaction length scale for a spherical microswimmer and recover the equations discussed in \cite{stone1996propulsion}. 

We proceed to simplify Eqs. \eqref{Eq: unified_trans_exp_for_spheres}-\eqref{Eq: unified_rot_exp_for_spheres} at the thin interaction length limit, ${\lambda}/{a} \ll 1$, where $\lambda$ is the interaction lengthscale. To this end, we define a stretched radial coordinate $\rho = {(r-a)}/{\lambda}$. Next, we expand and re-write Eqs. \eqref{Eq: unified_trans_exp_for_spheres}-\eqref{Eq: unified_rot_exp_for_spheres} in orders of ${\lambda}/{a}$. Subsequently, the leading order contribution to the translation and rotation velocities are obtained to be,
\begin{equation}
    \mathbf{U} = -\frac{\lambda}{4\pi\mu a^2} \displaystyle \int_V \rho \mathbf{b}_\parallel\;{\rm dV}, \label{Eq: translation_asymptotics}
\end{equation}
\begin{equation}
    \mathbf{\Omega} = -\frac{3\lambda}{8\pi\mu a^3} \displaystyle \int_V \rho \mathbf{e}_r \times \mathbf{b}_\parallel\;{\rm dV} \label{Eq: rotation_asymptotics}.
\end{equation}
Since the volume of interest at the thin interaction limit is a spherical shell of thickness $\lambda$ surrounding the particle, we can rewrite the differential volume element to be, ${\rm dV} = \lambda{\rm d\rho dS}$ and the volume integrals as,
\begin{equation}
    \mathbf{U} = -\frac{\lambda^2}{4\pi\mu a^2} \displaystyle \int_{S_p}\left[\displaystyle \int_0^\infty \rho \mathbf{b}_\parallel\;{\rm d \rho}\right]\;{\rm dS}, \label{Eq: translation_volume_integral_decomposition}
\end{equation}
\begin{equation}
    \Omega = -\frac{3\lambda^2}{8\pi\mu a^3} \displaystyle \int_{S_P} \mathbf{e}_r \times \left[\displaystyle \int_0^\infty \rho \mathbf{b}_\parallel\;{\rm d}\rho\right]\;{\rm d}S. \label{Eq: rotation_volume_integral_decomposition}
\end{equation}
In the thin interaction limit, the shear force is balanced by the parallel body force, or
\begin{equation}
    \frac{\mu}{\lambda^2}\frac{\partial^2 \mathbf{u}_\parallel}{\partial \rho^2} + \mathbf{b}_\parallel = 0 \label{Eq: tangential_force_balance}.
\end{equation}
We note that in Eq. \eqref{Eq: tangential_force_balance}, since $\mathbf{b}$ is the osmophoretic force,  $\mathbf{b}_\parallel$ also includes the excess osmotic term. Multiplying Eq. \eqref{Eq: tangential_force_balance} by $\rho$ and employing integration by parts, we arrive at
\begin{equation}
    \displaystyle \int_0^\infty \rho \mathbf{b}_\parallel\;{\rm d\rho} = \frac{\mu}{\lambda^2} \mathbf{u}_\parallel |_0^\infty = \frac{\mu}{\lambda^2} \mathbf{u}_s, \label{Eq: integral_tangential_force_balance}
\end{equation}
\noindent where $\mathbf{u}_s = \mathbf{u}_{\parallel,\infty}-\mathbf{u}_{\parallel,0}$, is the phoretic slip velocity. Substituting Eq. \eqref{Eq: integral_tangential_force_balance} into Eqs. \eqref{Eq: translation_volume_integral_decomposition}-\eqref{Eq: rotation_volume_integral_decomposition}, we get the widely used result derived in \cite{stone1996propulsion},
\begin{equation}
    \mathbf{U} = -\frac{1}{4\pi a^2} \displaystyle \int_{S_P} \mathbf{u}_s\;{\rm dS}, \label{Eq: stone and samuel translation}
\end{equation}
\begin{equation}
    \mathbf{\Omega} = -\frac{3}{8\pi a^3} \displaystyle \int_{S_P} \mathbf{e}_r \times \mathbf{u}_s\; {\rm dS}, \label{Eq: stone and samuel rotation}
\end{equation}
where the integral is over the surface of the sphere. 
\subsection{Electrophoretic mobility at arbitrary interaction lengthscales} \label{ssec: electrophoresis}

To further validate Eqs. \eqref{Eq: unified_trans_exp_for_spheres}-\eqref{Eq: unified_rot_exp_for_spheres} by the determination of the electrophoretic mobility of a sphere in the Debye-H\"{u}ckel limit for an arbitrary Debye length \citep{henry1931cataphoresis,teubner1982motion,kim2013microhydrodynamics}. We assume a homogeneous sphere of radius, $a$, immersed in a binary monovalent electrolytic solution such that the electrical permittivity of the solution is denoted as $\varepsilon$. Our objective is to analyze the motion of the particle with a given surface zeta potential driven by an external electric field. First, we assume that the surface zeta potential, $\zeta$, falls in the Debye-H\"{u}ckel limit, ${e\zeta}/{k_B T} \ll 1$, where $e$ is the charge of an electron, $k_B$ is the Boltzmann constant, and $T$ is the absolute temperature. We also assume an electric field disturbance of $\mathbf{E}_\infty$ far away from the particle such that $\mathbf{E}_\infty = \epsilon E_0 \mathbf{e}_z$, where $\epsilon$ is a small parameter and physically indicates that the length scale of far-field potential decay is much larger than the particle size. Note that $\varepsilon$ is the electrical permittivity and should not be confused with $\epsilon$, which is a small parameter in our analysis. Finally, the total osmophoretic body force ($\mathbf{b}$) driving the particle arises through a combination of the electrostatic interaction and net excess osmotic pressure in the fluid, or 
\begin{figure}
    \centering
    \includegraphics[width=\linewidth]{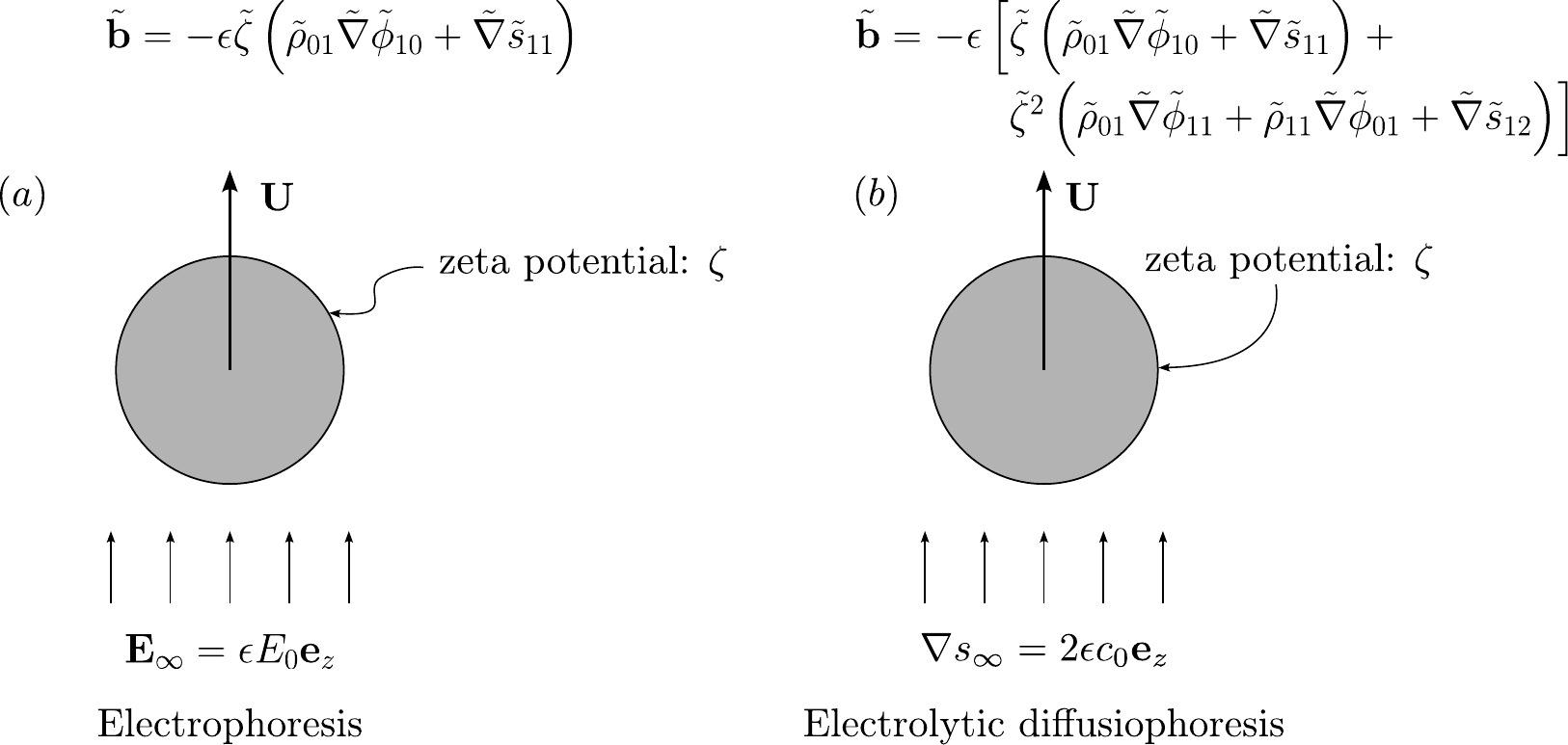}
    \caption{Methodology to validate proposed mobility expressions for a charged particle with a zeta potential ($\zeta$) in the Debye-H\"{u}ckel limit for (a) electrophoresis with an external field $\mathbf{E}_\infty=\epsilon E_0 \mathbf{e}_z$ and (b) diffusiophoresis with externally imposed solute gradient $\bm{\nabla}s_\infty = 2\epsilon c_0 \mathbf{e}_z$. The expressions of dimensionless osmophoretic force $\tilde{\mathbf{b}}$ are provided. Substituting the appropriate $\mathbf{b}$ in Eq. \eqref{Eq: unified_trans_exp_for_spheres} enables us to recover mobility relationships that otherwise require cumbersome calculations.}
    \label{fig:electrophoretic_mobility_validation}
\end{figure}
\begin{equation}
    \mathbf{b} = -e\left(c_{+} - c_{-}\right) \bm{\nabla} \phi - k_B T \bm{\nabla}\left(c_{+} + c_{-}\right), \label{Eq: electrophoretic_body_force}
\end{equation}

\noindent where $c_{+}$ and $c_{-}$ are the concentrations of the positive and negative electrolytic species respectively and $\phi$ is the electric potential. As a convenient choice, we can represent the solute concentrations in terms of net charge, $\rho = e \left( c_{+} - c_{-} \right)$, and salt, $s = c_{+} + c_{-}$, and rewrite the body force to be $\mathbf{b} = - \rho \bm{\nabla} \phi - k_B T \bm{\nabla} s$. It should be noted that care should be exercised in choosing the appropriate expression for $\mathbf{b}$. Specifically, the osmotic contribution $-k_B T \bm{\nabla}s$ refers to the excess osmotic contribution arising out of an interaction that locally drives the solute out of equilibrium. This effectively implies that in the absence of such interactions, an external salt gradient on its own cannot induce net particle motion, as demonstrated in \cite{brady2021phoretic}. The equivalence of Eq. \eqref{Eq: unified_trans_exp_for_spheres} and the results of \cite{brady2021phoretic} can be seen by defining an additional surface stress contribution, $\bm{\sigma}_p = -k_B T s \mathbf{I}$, as per Eq. (2.17) in \cite{brady2021phoretic}, due to the excess osmotic pressure. The divergence of $\bm{\sigma}_p$ leads to the second term in Eq. \eqref{Eq: electrophoretic_body_force}, $-k_B T \bm{\nabla} s$. We refer the readers to our discussion on the mechanistic origin of $\mathbf{b}$ in section \ref{sec:deriv} and redirect them to \cite{brady2021phoretic} for more details.

To appropriately derive the particle motion, we are required to obtain the solutions to $\rho$, $s$, and $\phi$ for a given $\mathbf{E}_\infty$ and $\zeta$. As mentioned earlier, we assume that the species are monovalent, $z_\pm=\pm1$ and have  diffusivities $D_\pm$, the species balance is given by the steady Nernst-Planck equations,
\begin{equation}
    \mathbf{u}\cdot\bm{\nabla}c_+ =D_+ \left[\nabla^2 c_{+} + \frac{e}{k_B T} \bm{\nabla}\cdot\left(c_{+}\bm{\nabla}\phi\right)\right] \label{Eq: positive_species_balance}
\end{equation}
\begin{equation}
     \mathbf{u}\cdot\bm{\nabla}c_- = D_- \left[\nabla^2 c_{-} - \frac{e}{k_B T} \bm{\nabla}\cdot\left(c_{-}\bm{\nabla}\phi\right)\right]. \label{Eq: negative_species_balance}
\end{equation}
For $\mathrm{Pe} = aU/D \ll 1$ ($U$ is the velocity scale for the particle, $D = \frac{2 D_+ D_-}{D_+ + D_-}$ is the ambipolar diffusivity), we ignore the convective effects. Consequently, Eqs. \eqref{Eq: positive_species_balance} and \eqref{Eq: negative_species_balance} can be rewritten in terms of $\rho$ and $s$ as
\begin{equation}
    \nabla^2 s + \frac{1}{k_B T} \bm{\nabla}\cdot\left(\rho\bm{\nabla}\phi\right) = 0, \label{Eq: salt_balance}
\end{equation}
\begin{equation}
    \nabla^2\rho + \frac{e^2}{k_B T} \bm{\nabla}\cdot\left(s\bm{\nabla}\phi\right) = 0. \label{Eq: charge_balance}
\end{equation}
Finally, the system of equations is closed by using Poisson's equation to resolve the electric potential,
\begin{equation}
   - \varepsilon \nabla^2 \phi = \rho. \label{Eq: poissons_equation}
\end{equation}
In the far-field, at $r\to\infty$, the potential gradient is the externally imposed electric field and the fluid is electroneutral, or 
\begin{equation}
   \left. -\bm{\nabla}\phi  \right|
_{r \rightarrow \infty}= \epsilon E_0 \mathbf{e}_z. \label{Eq: bc_potential_far-field}
\end{equation}
\begin{equation}
\left.    \rho \right|_{r \rightarrow \infty} = 0. \label{Eq: bc_charge_far-field}
\end{equation}
Moreover, in the far-field $c_{+} = c_{-} = c_0$, where $c_0$ is a characteristic solute concentration, we can write $s$ to follow 
\begin{equation}
    \left. s \right|_{r \rightarrow \infty} = 2c_0  
\end{equation}
At the particle surface, at $r=a$, the electrostatic potential is equal to the zeta potential at the surface, or
\begin{equation}
    \left. \phi \right|_{r = a} = \zeta. \label{Eq: bc_surface_potential}
\end{equation}
Additionally, there is no salt or charge flux normal to the particle surface,

\begin{equation}
    \mathbf{e}_r\cdot\left[\bm{\nabla}s + \frac{\rho}{k_B T} \bm{\nabla}\phi\right]_{r = a} = 0, \label{Eq: bc_surface_salt}
\end{equation}
\begin{equation}
    \mathbf{e}_r\cdot\left[\bm{\nabla}\rho + \frac{s e^2}{k_B T} \bm{\nabla}\phi\right]_{r = a} = 0.\label{Eq: bc_surface_charge}
\end{equation}
\noindent Eqs. \eqref{Eq: salt_balance}-\eqref{Eq: bc_surface_charge} are non-dimensionalized using the following appropriate scales,
\begin{equation}
    \Tilde{\bm{\nabla}} = a \bm{\nabla},\;\;\;\Tilde{\nabla}^2 = a^2 \nabla^2,\;\;\;\Tilde{\phi} = \frac{e\phi}{k_B T},\;\;\;\Tilde{\rho} = \frac{\rho}{e c_0},\;\;\;\Tilde{s} = \frac{s}{c_0}, \;\;\;\Tilde{r} = \frac{r}{a}. \label{Eq: characteristic_scales}
\end{equation}
\noindent Thus the non-dimensional Poisson-Nernst-Planck equations are given as,
\begin{eqnarray}
    \Tilde{\nabla}^2\Tilde{s} + {\Tilde{\bm{\nabla}}}\cdot\left(\Tilde{\rho}\Tilde{\bm{\nabla}}\Tilde{\phi}\right) = 0, \label{Eq: non-dim-salt-bal} \\
    \Tilde{\nabla}^2\Tilde{\rho} + \Tilde{\bm{\nabla}}\cdot\left(\Tilde{s}\Tilde{\bm{\nabla}}\Tilde{\phi}\right) = 0, \label{Eq: non-dim-charge-bal} \\
    \Tilde{\nabla}^2 \Tilde{\phi} = -\frac{\kappa^2}{2} \Tilde{\rho}, \label{Eq: non-dim-poisson-eq}
\end{eqnarray}
where $\kappa = \left(\frac{2 a^2 e^2  c_0 }{\varepsilon k_B T}\right)^{1/2}$ is the dimensionless inverse Debye length. In the far field, thus
\begin{eqnarray}
    -\left. \Tilde{\bm{\nabla}}\Tilde{\phi} \right|_{\Tilde{r} \rightarrow \infty} = \epsilon \Tilde{E}_0 \mathbf{e}_z, \label{Eq: non-dim-far-field-potential-BC} \\
   \left. \Tilde{\rho} \right|_{\Tilde{r} \rightarrow \infty} = 0 , \label{Eq: non-dim-far-field-charge-BC} \\
   \left. \Tilde{s} \right|_{\Tilde{r} \rightarrow \infty} = 2  , \label{Eq: non-dim-far-field-salt-BC}
\end{eqnarray}
where $\Tilde{E}_0 = a e E_0/\left(k_B T \right)$ is the non-dimensional electric-field. Similarly, at the particle surface, the non-dimensional boundary conditions read 
\begin{eqnarray}
        \mathbf{e}_r \cdot \left[\Tilde{\bm{\nabla}}\Tilde{s} + \Tilde{\rho} \Tilde{\bm{\nabla}}\Tilde{\phi}\right]_{\Tilde{r} = 1} = 0, \label{Eq: non-dim-surface-salt-BC} \\
    \mathbf{e}_r \cdot \left[\Tilde{\bm{\nabla}}\Tilde{\rho} + \Tilde{s} \Tilde{\bm{\nabla}} \Tilde{\phi}\right]_{\Tilde{r} = 1} = 0, \label{Eq: non-dim-surface-charge-BC} \\
    \left. \Tilde{\phi} \right|_{\Tilde{r} = 1} = \frac{e \zeta}{k_B T} = \Tilde{\zeta}. \label{Eq: non-dim-surface-potential-BC}
\end{eqnarray}
As we discuss later, it is more appropriate to write Eq. \eqref{Eq: non-dim-surface-potential-BC} as a constant charge boundary condition, which renders the gradient of the potential to be constant instead. However, for the weak-field, these surface boundary conditions are equivalent and hence we retain the constant potential boundary condition for simplicity. 
\par{} For the remainder of the calculation until Eq. \eqref{Eq: ord-11-radial-dependence-salt-dynamics}, we will drop the tilde superscript in Eqs. \eqref{Eq: non-dim-salt-bal}-\eqref{Eq: non-dim-surface-potential-BC} for convenience and restore the dimensions once the non-dimensional calculations are complete. We expand $\phi$, $\rho$, and $s$ in the small parameters $\zeta$ and $\epsilon$ as
\begin{equation}
    \phi = \phi_{00} + \zeta \phi_{01} + \epsilon \left(\phi_{10} + \zeta\phi_{11}\right), \label{Eq: expansion_in_phi}
\end{equation}
\begin{equation}
    \rho = \rho_{00} + \zeta \rho_{01} + \epsilon\left(\rho_{10} + \zeta \rho_{11}\right), \label{Eq: expansion_in_rho}
\end{equation}
\begin{equation}
    s = s_{00} + \zeta s_{01} + \epsilon \left(s_{10} + \zeta s_{11}\right). \label{Eq: expansion_in_salt}
\end{equation}
The asymptotic expansions in Eqs.\eqref{Eq: expansion_in_phi}-\eqref{Eq: expansion_in_salt} are substituted into Eqs. \eqref{Eq: non-dim-salt-bal}-\eqref{Eq: non-dim-surface-potential-BC}, and the corresponding equations are solved at each asymptotic order. 

$\bm{O(1)}$: \textbf{An uncharged particle without any electric field.} 
\noindent The governing equations and boundary conditions are obtained to be
\begin{eqnarray}
        \nabla^2 s_{00} + \bm{\nabla}\cdot\left(\rho_{00} \bm{\nabla}\phi_{00}\right) = 0, \label{Eq: lead-ord-salt} \\
    \nabla^2 \rho_{00} + \bm{\nabla}\cdot\left(s_{00}\bm{\nabla}\phi_{00}\right) = 0, \label{Eq: lead-ord-charge} \\
    \nabla^2 \phi_{00} = -\frac{\kappa^2}{2} \rho_{00}. \label{Eq: lead-ord-poisson} \\
\mathbf{e}_r\cdot\left[\bm{\nabla}s_{00}+\rho_{00}\bm{\nabla}\phi_{00}\right]=0,\;\;\;\mathrm{at}\;\;r=1, \label{Eq: lead-ord-surface-salt-flux} \\
    \mathbf{e}_r \cdot \left[\bm{\nabla} \rho_{00} + s_{00} \bm{\nabla}\phi_{00}\right]=0,\;\;\;\mathrm{at}\;\;r=1, \label{Eq: lead-ord-surface-charge-flux} \\
    \phi_{00} = 0,\;\;\;\mathrm{at}\;\;r=1. \label{Eq: lead-ord-surface-potential} \\    s_{00}=2,\;\;\;\mathrm{at}\;\;r\to\infty, \label{Eq: lead-ord-far-fld-salt} \\
\rho_{00}=0,\;\;\;\mathrm{at}\;\;r\to\infty, \label{Eq: lead-ord-far-fld-charge} \\    \bm{\nabla}\phi_{00}=0,\;\;\;\mathrm{at}\;\;r\to\infty, \label{Eq: lead-ord-far-fld-potential}
\end{eqnarray}

The system of Eqs. \eqref{Eq: lead-ord-salt}-\eqref{Eq: lead-ord-far-fld-salt} have a trivial solution, \textit{i.e.}, $\phi_{00}=0$, $\rho_{00}=0$, $s_{00}=2$. Physically, the solution simply implies that the ion concentration is uniform because the particle is uncharged and there is no electric field. 

$\bm{O(\epsilon)}$: \textbf{Perturbation due to the electric field for an uncharged particle.} The governing equations for the salt and charge dynamics, and electrostatic potential after substituting the expressions of $\rho_{00}$, $s_{00}$, and $\phi_{00}$ are given as,
\begin{equation}
    \nabla^2 s_{10}=0, \label{Eq: ord-10-salt-reduced}
\end{equation}
\begin{equation}
    \nabla^2 \rho_{10} + 2 \nabla^2 \phi_{10}=0, \label{Eq: ord-10-charge-balance-reduced}
\end{equation}
\begin{equation}
    \nabla^2 \phi_{10} = -\frac{\kappa^2}{2} \rho_{10}. \label{Eq: ord-10-poisson-reduced}
\end{equation}
The corresponding reduced boundary conditions when $r \to \infty$ are
\begin{equation}
    s_{10} = 0, \label{Eq: ord-10-far-field-salt-reduced}
\end{equation}
\begin{equation}
    \rho_{10}=0, \label{Eq: ord-10-far-field-charge-reduced}
\end{equation}
\begin{equation}
    -\bm{\nabla}\phi_{10} = E_0 \mathbf{e}_z, \label{Eq: ord-10-far-field-potential-reduced}
\end{equation}
and when $r=1$, they read
\begin{equation}
    \mathbf{e}_r\cdot\bm{\nabla}s_{10}=0, \label{Eq: ord-10-surface-salt-reduced}
\end{equation}
\begin{equation}
    \mathbf{e}_r\cdot\left[\bm{\nabla}\rho_{10}+2\bm{\nabla}\phi_{10}\right]=0. \label{Eq: ord-10-surface-charge-reduced}
\end{equation}
\begin{equation}
    \mathbf{e}_r \cdot \bm{\nabla} \phi_{10}=0, \label{Eq: ord-10-surface-potential-reduced}
\end{equation}
\noindent where the derivative of the potential is set to be zero to ensure that there is no excess charge on the surface; see discussion below Eq. \eqref{Eq: non-dim-surface-potential-BC}. Mathematically, at $O(\epsilon)$, the charge on the particle surface is zero. Through Gauss's law, the no surface charge boundary condition necessitates that $\mathbf{e}_r \cdot\bm{\nabla}\phi_{10} = 0$. Hence, Eq. \eqref{Eq: ord-10-surface-charge-reduced} implies that  $\mathbf{e}_r\cdot\bm{\nabla}\rho_{10}=0$ at the particle surface. Along with Eqs. \eqref{Eq: ord-10-charge-balance-reduced} and \eqref{Eq: ord-10-far-field-charge-reduced},  we obtain $\rho_{10}=0$. Similarly, Eqs. \eqref{Eq: ord-10-salt-reduced}, \eqref{Eq: ord-10-far-field-salt-reduced} and \eqref{Eq: ord-10-surface-salt-reduced} reveal $s_{10}=0$. 
\par{} Since $\rho_{10}=0$, $\phi_{10}$ is governed by the Laplace equation, and the solution with appropriate boundary conditions reads \citep{griffiths2005introduction},
\begin{equation}
    \phi_{10}\left(r,\theta\right) = -E_0 r \left(1+\frac{1}{2r^3}\right) \cos \theta. \label{Eq: ord-10-potential-solution}
\end{equation} 
Physically, this order implies that perturbed potential and corresponding electric field lines get modified due to the geometry of the particle but there is no charge and salt accumulation. 

$\bm{O(\zeta)}$: \textbf{Perturbation of a charged particle without an external electric field.} The equations governing the charge, salt, and potential at $O(\zeta)$ are analogous to Eqs. \eqref{Eq: ord-10-salt-reduced}-\eqref{Eq: ord-10-poisson-reduced} at $O(\epsilon)$ but have different boundary conditions. The governing equations are
\begin{eqnarray}
    \nabla^2 s_{01} = 0, \label{Eq: ord-01-salt-reduced} \\
    \nabla^2 \rho_{01} + 2 \nabla^2 \phi_{01} =0, \\
    \nabla^2 \phi_{01} = -\frac{\kappa^2}{2}\rho_{01}. \label{Eq: ord-01-poisson-reduced}
\end{eqnarray}
As $r \to \infty$ the net charge, salt, and electric potential gradient all decay to zero, or
\begin{eqnarray}
    s_{01} = 0, \label{Eq: ord-01-far-fld-salt} \\
    \rho_{01} = 0, \label{Eq: ord-01-far-fld-charge} \\
    \bm{\nabla} \phi_{01} = 0. \label{Eq: ord-01-far-fld-potential}
\end{eqnarray}
At the particle surface, $r=1$, we obtain
\begin{eqnarray}
\mathbf{e}_r\cdot\bm{\nabla}s_{01} = 0, \label{Eq: ord-01-surface-salt} \\ \mathbf{e}_r\cdot\left[\bm{\nabla}\rho_{01} + 2\bm{\nabla}\phi_{01}\right]=0, \label{Eq: ord-01-surface-charge} \\
    \phi_{01}=1. \label{Eq: ord-01-surface-potential}
\end{eqnarray}
Eqs. \eqref{Eq: ord-01-salt-reduced}, \eqref{Eq: ord-01-far-fld-salt}, and \eqref{Eq: ord-01-surface-salt} yield $s_{01} = 0$. The remainder of the equations reveal 
\begin{equation}
    \phi_{01}(r) = \frac{1}{r} e^{-\kappa\left(r-1\right)} \label{Eq: ord-01-potential-sol}
\end{equation}
\begin{equation}
    \rho_{01}(r) = -\left(\frac{2}{r}\right)e^{-\kappa\left(r-1\right)} \label{Eq: ord-01-charge-sol}
\end{equation}
Physically, the results indicate the distribution of potential and charge arising due to a charged particle.

$\bm{O(\epsilon\zeta)}$: \textbf{Perturbation of both imposed electric field and surface zeta potential.} To fully resolve the body force to the order of $O(\epsilon\zeta)$,  we have to obtain $s_{11}$, this can be observed by the expansion of Eq. \eqref{Eq: electrophoretic_body_force} and collecting the respective orders. The governing equations and boundary conditions for salt, $s_{11}$, are given by,
\begin{equation}
    \nabla^2 s_{11} + \bm{\nabla}\cdot\left(\rho_{01}\bm{\nabla} \phi_{10}\right) = 0, \label{Eq: ord-11-salt-dynamics}
\end{equation}
\begin{equation}
    s_{11} \to 0,\;\;\;\mathrm{at}\;r \to \infty, \label{Eq: ord-11-far-fld-slt}
\end{equation}
\begin{equation}
    \mathbf{e}_r \cdot \left[\bm{\nabla}s_{11} + \rho_{01}\bm{\nabla}\phi_{10}\right]=0\;\;\;\mathrm{at}\;r=1. \label{Eq: ord-11-surf-salt}
\end{equation}
Substituting in the expression of $\rho_{01}$ from Eq. \eqref{Eq: ord-01-charge-sol} and $\phi_{10}$ from Eq. \eqref{Eq: ord-10-potential-solution} we can solve for $s_{11}(r,\theta)$. The salt dynamics are in the form of $s_{11}=E_0 f(r)\cos \theta$ where $f(r)$ is
\begin{equation}
    \begin{aligned}
        f(r) = &\frac{1}{r^2}\left(\frac{\kappa}{6}-\frac{5}{3}-\frac{10}{\kappa}-\frac{2}{\kappa^2}\right)+\frac{\kappa^2}{3}e^{\kappa} \textrm{Ei} \left(-\kappa r\right)+\frac{2}{\kappa^2 r^2} e^{-\kappa\left(r-1\right)}-\\&\frac{4}{3\kappa r^2}e^{-\kappa(r-1)}+\frac{ \kappa}{3} e^{-\kappa(r-1)}-\frac{1}{3r}e^{-\kappa(r-1)}+\frac{2}{\kappa r}e^{-\kappa(r-1)}, \label{Eq: ord-11-radial-dependence-salt-dynamics}
    \end{aligned}
\end{equation}
\noindent where $\textrm{Ei}()$ is the elliptic integral.  \\
\textbf{Reintroducing dimensions.} At this stage, we restore the dimensions and reintroduce the tilde for dimensionless variables. The total osmophoretic body force $\mathbf{b}$ is made dimensionless by writing $\mathbf{b} = \frac{k_B T c_0}{a} \tilde{ \mathbf{b} } = \frac{ \varepsilon \left(k_B T \right)^2  \kappa^2 }{2 e^2 a^3} \tilde{\mathbf{b}}$. The first relevant order of $\tilde{\mathbf{b}}$ for electrophoresis is $\epsilon \tilde{\zeta}$ because it is the order at which a charged particle is being driven by an electric field. To this end, we write 
\begin{align}
    \begin{split}
        \tilde{\mathbf{b}}  = 
    - \epsilon \tilde{\zeta} \left[ \tilde{\rho}_{00} \tilde{\bm{\nabla}} \tilde{\phi}_{11}  
    + \tilde{\rho}_{11} \tilde{\bm{\nabla}} \tilde{\phi}_{00} + \tilde{\rho}_{01} \tilde{\bm{\nabla}} \tilde{\phi}_{10} + \tilde{\rho}_{10} \tilde{\bm{\nabla}} \tilde{\phi}_{01} + \tilde{\bm{\nabla}} \tilde{s}_{11} \right].  \label{Eq: electrophoretic_body_force_asymptotic_def}
    \end{split}
\end{align}
Based on the solutions at different orders, it is straightforward to see that $\mathbf{\tilde{b}}$ reduces to 
\begin{equation}
    \mathbf{\tilde{b}} = -\epsilon \tilde{\zeta} \left[ \tilde{\rho}_{01} \Tilde{\bm{\nabla}} \tilde{\phi}_{10} + \Tilde{\bm{\nabla}} \tilde{s}_{11} \right].
\end{equation}
We note that the body force term of  $\tilde{s}_{11}$ integrates out to zero in the calculation of $\mathbf{U}$ and $\mathbf{\Omega}$ for electrophoresis and is generally not included in prior analyses. However, we retain this term for consistency as it does become crucial for diffusiophoretic phenomena, as we detail in section \ref{ssec: diffusiophoresis}. 

After substituting the values of $\tilde{\rho}_{01}$, $\tilde{\phi}_{10}$ and $\tilde{s}_{11}$, the resultant equation in dimensional form is
\begin{equation}
    \begin{aligned}
    \mathbf{b}=-\frac{\epsilon \varepsilon E_0 \zeta \kappa^2} {2 a^2} &\left\{\left[\frac{2}{\tilde{r}}e^{-\kappa(\tilde{r}-1)}\left(1-\frac{1}{\tilde{r}^3}\right)\cos\theta+\frac{d f(\tilde{r})}{d \tilde{r}}\cos\theta\right]\mathbf{e}_r \right. \\ &\left .-\left[\frac{2}{\tilde{r}}e^{-\kappa(\tilde{r}-1)} \left(1+\frac{1}{2 \tilde{r}^3}\right)\sin\theta+\frac{f(\tilde{r})}{\tilde{r}}\sin \theta\right]\mathbf{e}_\theta \right\}. \label{Eq: ord-11-electrophoretic-body force}
    \end{aligned}
\end{equation}
\noindent Eq. \eqref{Eq: ord-11-electrophoretic-body force} is substituted into Eq. \eqref{Eq: unified_trans_exp_for_spheres} to obtain the translational velocity to be
\begin{equation}
    \mathbf{U}=\mathcal{M} \mathbf{E}_{\infty}, \label{Eq: electrophoretic_mobility_relation}
\end{equation}
where the mobility $\mathcal{M}$ after restoring dimensions is
\begin{equation}
\mathcal{M}=\frac{\varepsilon\zeta}{6\mu}\left[(1+\kappa)+(12-\kappa^2)\displaystyle \int_1^\infty \frac{e^{\kappa (1-t)}}{t^5} {\rm d}t\right], \label{Eq: electrophoretic_mobility_expression}
\end{equation}
which is the seminal result of \cite{henry1931cataphoresis} for arbitrary double layer thickness, and has also been reported by \cite{teubner1982motion} and \cite{kim2013microhydrodynamics}. For a homogeneous sphere, our analysis reveals $\mathbf{\Omega}=\mathbf{0}$, as expected. We emphasize that it is straightforward to extend the calculations to heterogeneous spheres \citep{velegol1996probing, teubner1982motion} and obtain results for electrorotation in arbitrary double-layer thicknesses, which otherwise requires considerable effort. 

\subsection{Electrolytic diffusiophoretic mobility at arbitrary interaction lengthscales}\label{ssec: diffusiophoresis}
Next, we focus on the process of electrolytic diffusiophoresis in the Debye-H\"{u}ckel limit and for arbitrary double-layer thickness. We assume that the external concentration gradient of a binary monovalent electrolyte is given as $\bm{\nabla} s_{\infty} = 2 \epsilon \bm{\nabla} c_0 $, where $\epsilon$ is a small parameter, much like electrophoresis. Here, $\mathbf{b}$ is required to be expanded to an additional higher order of  $\epsilon \tilde{\zeta}^2$. The term on the order $O \left( \epsilon \tilde{\zeta} \right)$ is identical to electrophoresis and represents the electrophoretic component of the diffusiophoretic mobility. The second term on the order O $\left( \epsilon \tilde{\zeta}^2 \right)$ denotes the chemiphoretic component.
We employ the expression of $\mathbf{b}$, derived in this sub-section, to Eqs. \eqref{Eq: unified_trans_exp_for_spheres}-\eqref{Eq: unified_rot_exp_for_spheres}. This allows us to retrieve the expression of the translation velocity of a charged spherical particle in an unbounded solution of a symmetrically charged electrolyte for an arbitrary double-layer thickness, which otherwise requires considerable efforts; see \cite{keh2000diffusiophoretic}. 

We acknowledge the electrokinetic equations used to describe such diffusiophoretic systems are analogous to our treatment of the motion of electrophoretically propelled particles in section \ref{ssec: electrophoresis}. However, the key mechanistic difference is the presence of an external gradient of solute $\bm{\nabla} s_{\infty}$ instead of an imposed electric field $\mathbf{E}_\infty$; this results in a change of boundary conditions and subsequently the solutions at different asymptotic orders. To preserve the pedagogical nature of our manuscript, we will re-derive the electrophoretic contribution and subsequently solve for the chemiphoretic contribution to the osmophoretic body force term and attempt to emphasize key physical and mathematical differences between the derivations laid out in sections \ref{ssec: electrophoresis} and \ref{ssec: diffusiophoresis}.

Consider a colloidal particle with a surface zeta potential, $\zeta$, in an external solute gradient of a symmetric binary electrolyte. We assume that the electrolytes are monovalent such that $z_\pm = \pm 1$. The ions are assumed to have different diffusivities $D_+ \neq D_-$. The governing equations of the concentration of the ionic species, $c_\pm$, are identical to Eqs. \eqref{Eq: positive_species_balance}-\eqref{Eq: negative_species_balance} and the interaction potential is governed by Poisson's equation, as given in Eq. \eqref{Eq: poissons_equation}. However, the far-field boundary conditions are different. Specifically, as $r \to \infty$, we assume that the concentration of the ionic species is linear with position $z$, or
\begin{equation}
    c_\pm = c_0\left(1+ \frac{\epsilon z}{a}\right), \label{Eq: diff_bc_far_fld_cation_conc}
\end{equation}
Further, it is assumed that the electric current in the far field is zero, which yields \citep{prieve1984motion, velegol2016origins, gupta2019diffusiophoretic}
\begin{equation}
-  \bm{\nabla} \phi_{\infty} = \epsilon \beta \frac{k_B T}{a e} = \epsilon E_0 \mathbf{e}_z, \label{Eq: no_current_boundary_condition}
\end{equation}
where $\beta = \frac{D_+ - D_-}{D_+ + D_-}$ and $E_0 = \beta \frac{k_B T}{a e}$. The electric field is thus induced due to unequal diffusivities and a non-zero salt gradient. 

Similar to electrophoresis, modification of the governing equations in terms of charge, $\rho = e\left(c_+ - c_-\right)$, salt, $s=c_+ + c_-$, and potential result in Eqs. \eqref{Eq: salt_balance}-\eqref{Eq: poissons_equation}. In the far field, the boundary conditions for $\phi$ and $\rho$ are identical to Eqs. \eqref{Eq: bc_potential_far-field} and \eqref{Eq: bc_charge_far-field} with the aforementioned definition of $E_0$. However, the boundary condition of $s$ is modified to
\begin{equation}
    s|_{r \to\infty } = 2c_0 \left( 1+\frac{\epsilon z}{a} \right). \label{Eq: diff_bc_far_fld_salt}
\end{equation}
Note that $\epsilon/a = \bm{\nabla} \log s_\infty$. The boundary conditions are identical at the particle surface, i.e., Eqs. \eqref{Eq: bc_surface_potential}-\eqref{Eq: bc_surface_charge}. The objective is to solve $\rho$, $s$, and $\phi$ with the modified boundary conditions above and subsequently evaluate the total osmophoretic body force, following the procedure used to obtain  Eq. \eqref{Eq: electrophoretic_body_force}. We non-dimensionalize the equations using the same scales as Eq. \eqref{Eq: characteristic_scales} and also define $\mathbf{b} = \frac{k_B T c_0}{a} \tilde{\mathbf{b}} = \frac{\varepsilon \left(k_B T\right)^2 \kappa^2}{2 e^2 a^3} \tilde{\mathbf{b}} $, where the definition of $\kappa$ is also identical. 

For simplicity, we drop the tilde from our analysis until Eq. \eqref{Eq: diffusiophoretic mobility} and reintroduce them afterward. Thus the non-dimensional osmophoretic body force is given as $\mathbf{b} = -\rho \bm{\nabla}\phi-\bm{\nabla}s$. We expand $\rho$, $s$ and $\phi$ until $O(\epsilon \zeta^2)$, and solve the equations at each order.

{$\bm{O(1)}$: \textbf{An uncharged particle without any external salt gradient.} } The results at this order are identical to electrophoresis and thus yield $s_{00}=2$, $\rho_{00}=0$, and $\phi_{00}=0$, indicating a uniform concentration of ion with no charge and potential. 

{$\bm{O(\epsilon)}$: \textbf{Perturbation of the external salt concentration to an uncharged particle.}} This order is distinct compared to electrophoresis since the far-field boundary condition for salt is different, while the remainder of the equations and boundary conditions are identical. We note that the boundary condition for the electric field is similar to electrophoresis since we have defined $E_0$. The solution simply reduces to zero and uniform charge density $\rho_{10}=0$, while both salt and potential follow the Laplace equation. The results read

\begin{eqnarray}
    s_{10}\left(r,\theta\right)=2r\left(1+\frac{1}{2r^3}\right)\cos\theta. \label{Eq: diff_10_salt_solution} \\ 
        \phi_{10} = -E_0 r \left(1+\frac{1}{2r^3}\right)\cos\theta. \label{Eq: diff_10_potential_solution}
\end{eqnarray}
\noindent Physically, at this order, a gradient in the salt concentration far away perturbs the salt field and induces a potential field if diffusivity asymmetry is present ($E_0 \neq 0$ only when $\beta \neq 0$). However, since the surface is uncharged, $\rho_{10}=0$. 

{$\bm{O(\zeta)}$: \textbf{Perturbation in the surface charge of the particle without an external field.}} Since there is no external field at this order, the solution is identical to electrophoresis with $s_{01}=0$ and 
\begin{equation}
    \phi_{01}(r) = \frac{1}{r}e^{-\kappa(r-1)}, \label{Eq: ord_01_potential_solution}
\end{equation}
\begin{equation}
    \rho_{01}(r) = -\frac{2}{r} e^{-\kappa(r-1)}. \label{Eq: ord_01_charge_solution}
\end{equation}
\noindent This order represents the potential and charge profiles due to the surface charge of the particle. However, there is no salt accumulation at this order since the reduction in the co-ion concentration is balanced by the increase in the counter-ion concentration.

\textbf{$\bm{O(\epsilon\zeta)}$: Perturbation in both the imposed salt concentration and surface charge.} The governing equations for charge ($\rho_{11}$), salt ($s_{11}$), and potential ($\phi_{11}$) are
\begin{eqnarray}    
\nabla^2 s_{11} + \bm{\nabla}\cdot\left[\rho_{01}\bm{\nabla}\phi_{10} \right]=0, \label{Eq: diff_ord_11_salt_bal} \\
    \nabla^2 \rho_{11}+\bm{\nabla}\cdot\left[s_{10}\bm{\nabla}\phi_{01}+ 2 \bm{\nabla}\phi_{11}\right]=0, \label{Eq: diff_ord_11_chrg_bal} \\
    \nabla^2 \phi_{11} = -\frac{\kappa^2}{2}\rho_{11}. \label{Eq: diff_ord_11_potential_bal}
\end{eqnarray}
The boundary conditions at the particle surface, $r=1$, are
\begin{eqnarray}
\mathbf{e}_r\cdot\left[\bm{\nabla}s_{11}+\rho_{01}\bm{\nabla}\phi_{10}\right]=0, \label{Eq: diff_ord_11_surf_salt_flx} \\    \mathbf{e}_r\cdot\left[\bm{\nabla}\rho_{11}+s_{10}\bm{\nabla}\phi_{01}+2\bm{\nabla}\phi_{11}\right]=0, \label{Eq: diff_ord_11_surf_chrg_flx} \\
    \mathbf{e}_r\cdot \bm{\nabla} \phi_{11}=0. \label{Eq: diff_ord_11_surf_potential}
\end{eqnarray}

\noindent Again, there is no external field as $r\to\infty$.
Substituting in the solutions obtained in $O(\epsilon)$, and $O(\zeta)$ we begin to solve $s_{11}$, $\rho_{11}$, and $\phi_{11}$. Further, we can separate the $r$ and $\theta$ contributions by redefining $s_{11}\left(r, \theta\right) = f_{s_{11}}(r)\cos\theta$, $\rho_{11}\left(r,\theta\right)=f_{\rho_{11}}(r)\cos\theta$, and $\phi_{11}\left(r, \theta\right)=f_{\phi_{11}}(r)\cos \theta$ and solve the equations numerically; see Appendix A. It is possible to find analytical solutions to Eqs. \eqref{Eq: diff_ord_11_rho_radial_component}-\eqref{Eq: diff_ord_11_potential_flx_radial_component}, similar in form to Eq. \eqref{Eq: ord-11-radial-dependence-salt-dynamics}. In the scope of our current paper, we choose to resolve the dynamics at $O(\epsilon\zeta)$ and $O(\epsilon\zeta^2)$ numerically. For an analytical derivation of such higher-order effects the reader is directed to \cite{keh2000diffusiophoretic}.

{$\bm{O(\epsilon\zeta^2)}$: \textbf{First-order perturbation in salt field and second-order perturbation in surface charge}. } We only seek to solve $s_{12}$ at this order since it is the only quantity required to resolve the body force up to $O(\epsilon\zeta^2)$; see Eq. \eqref{Eq: diffusiophoretic_phoretic_body_force_12}.  The equation governing the dynamics of $s_{12}$ is,
\begin{equation}
    \nabla^2 s_{12} + \bm{\nabla}\cdot\left[\rho_{01}\bm{\nabla}\phi_{11}+\rho_{11}\bm{\nabla}\phi_{01}\right]=0, \label{Eq: diff_ord_12_salt_bal}
\end{equation}
with 
\begin{equation}
\mathbf{e}_r\cdot\left[\bm{\nabla}s_{12}+\rho_{01}\bm{\nabla}\phi_{11}+\rho_{11}\bm{\nabla}\phi_{01}\right]_{r=1}=0, \label{Eq: diff_ord_12_surf_salt_flx}
\end{equation}
and $s_{12}(r \rightarrow \infty, \theta) =0$. We write $s_{12}$ as $s_{12}\left(r,\theta\right)=f_{s_{12}}(r)\cos\theta$ and solve the equations numerically; see Appendix A. As discussed previously, we solve Eqs. \eqref{Eq: diff_ord_12_radial_salt_bal}, \eqref{Eq: diff_ord_12_surf_radial_salt_flx}, and the far field constraint numerically. 

\textbf{Restoring dimensions}. We now restore dimensions and reintroduce tilde to describe dimensionless variables. Therefore, we write the body force $\mathbf{b} = \frac{\varepsilon \left(k_B T\right)^2 \kappa^2}{2 e^2 a^3} \tilde{\mathbf{b}}$ such that 
\begin{eqnarray}
            \tilde{\mathbf{b}} = \epsilon \tilde{\zeta} \mathbf{\tilde{b}}_{11} + \epsilon \tilde{\zeta}^2 \mathbf{\tilde{b}}_{12}
            \label{Eq: diffusiophoretic_phoretic_body_force}
            \\
            \tilde{\mathbf{b}}_{11} =  - \left( \tilde{\rho}_{01} \bm{\tilde{\nabla}}\tilde{\phi}_{10}+\bm{\tilde{\nabla}}s_{11} \right) \\
           \tilde{\mathbf{b}}_{12} = - \left( \tilde{\rho}_{01}\bm{\tilde{\nabla}} \tilde{\phi}_{11}+\tilde{\rho}_{11}\bm{\tilde{\nabla}}\tilde{\phi}_{01}+\bm{\tilde{\nabla}}\tilde{s}_{12} \right)
\end{eqnarray}

At this point, some comments are in order. We note that $\tilde{ \mathbf{b}}$ could also include a term at the $O(\epsilon)$ since $s_{10} \neq 0$. However, $\tilde{\mathbf{b}}$ only includes excess osmotic pressure and not the osmotic pressure itself. This is because the osmotic pressure contribution due to $\bm{\nabla}s_\infty$ would lead to particle motion even in the absence of phoretic interactions. Only the terms at subsequent orders are included to ignore this effect. Further, we  highlight that both $\tilde{\phi}_{10}$  and $\tilde{s}_{11}$ are proportional to $\tilde{E}_0$. Therefore, the $O \left(\epsilon \tilde{\zeta} \right)$ term only depends on $E_0$ and is referred to as the electrophoretic contribution. Since $\tilde{\rho}_{01}$, $\tilde{\phi}_{10}$ and $\tilde{s}_{11}$ are identical to electrophoretic solution, the $O(\epsilon \tilde{\zeta})$ is equal to the one described earlier in Eq. \eqref{Eq: electrophoretic_mobility_expression}. 

\begin{figure}
    \centering
    \includegraphics[width=0.8\textwidth]{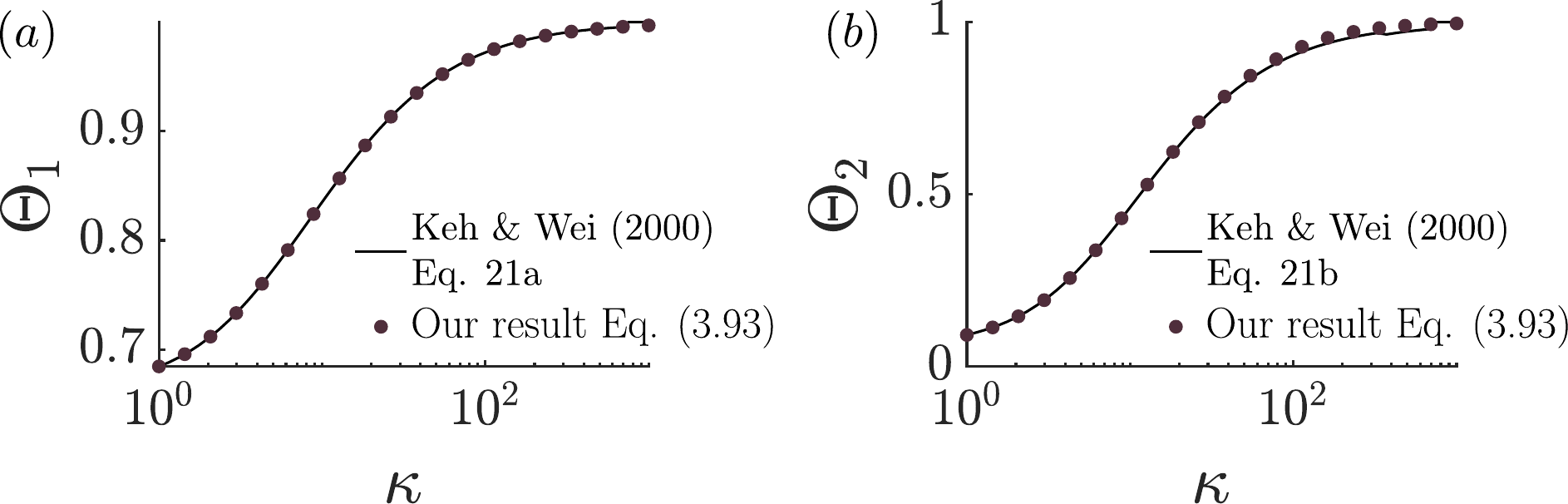}
    \caption{Comparison of proposed mobility expressions of diffusiophoretic mobility in Eq.~\eqref{Eq: diff_vel_keh} with the mobility reported in ~\cite{keh2000diffusiophoretic}. Quantitative agreement of both (a) $\Theta_1\left(\kappa\right)$ and (b) $\Theta_2\left(\kappa\right)$ is observed. }
    \label{fig:enter-label}
\end{figure}
However, in contrast, for the $O(\epsilon \tilde{\zeta}^2)$ contribution, $\tilde{\rho}_{01}$, $\tilde{\phi}_{11}$, $\tilde{\rho}_{11}$, $\tilde{\phi}_{01}$ and $\tilde{s}_{12}$ are independent of $E_0$. Furthermore, $\tilde{\phi}_{11}$, $\tilde{\rho}_{11}$ and $\tilde{s}_{12}$ are all proportional to $\tilde{s}_{10}$, which is consistent with the prior literature \citep{anderson1989colloid, keh2000diffusiophoretic, gupta2019diffusiophoretic}. 

Since $\tilde{\mathbf{b}}_{11}$ is identical to the electrophoretic motion, we focus our attention on $\tilde{\mathbf{b}}_{12}$, which reads

\begin{equation}
            \tilde{\mathbf{b}}_{12}=-  \left(\rho_{01}\frac{df_{\phi_{11}}}{dr}+f_{\rho_{11}}\frac{d\phi_{01}}{dr}+\frac{df_{s_{12}}}{dr} \right) \cos\theta \mathbf{e}_r -  \left(\frac{\rho_{01} f_{\phi_{11}}}{r}+\frac{f_{s_{12}}}{r}\right)\sin\theta \mathbf{e}_\theta \label{Eq: diffusiophoretic_phoretic_body_force_12}
\end{equation}

Upon substituting the value of $\mathbf{b}$ in Eq. \eqref{Eq: unified_trans_exp_for_spheres}, the translation velocity of the particle could be simplified to read
\begin{eqnarray}
    \mathbf{U} = \mathcal{M} \bm{\nabla} \log s_{\infty}, \label{Eq: diffusiophoretic mobility}
\end{eqnarray}
where 
\begin{equation}
    \mathcal{M} = \frac{\varepsilon}{\mu}\left[\frac{k_B T}{e}\beta \zeta  \Theta_1\left(\kappa\right) + \frac{\zeta^2}{8}\Theta_2\left(\kappa\right)\right], \label{Eq: diff_vel_keh}
\end{equation}
where $\Theta_1$ and $\Theta_2$ are evaluated numerically; see Appendix A. Figure \ref{fig:enter-label} demonstrates good quantitative agreement between the values obtained in \cite{keh2000diffusiophoretic} and our results. 

This section highlighted the generality of Eqs. \eqref{Eq: unified_trans_exp_for_spheres} and \eqref{Eq: unified_rot_exp_for_spheres}. We showed they are able to recover the mobilities for microswimmers in thin interaction layer limit, electrophoresis for arbitrary double-layer thickness, and electrolytic diffusiophoresis for arbitrary double-layer thickness.

\section{Autophoretic motion of microswimmers} \label{sec: autophoretic motion of microswimmers}
\begin{figure}
    \centering
    \includegraphics[width=0.9\textwidth]{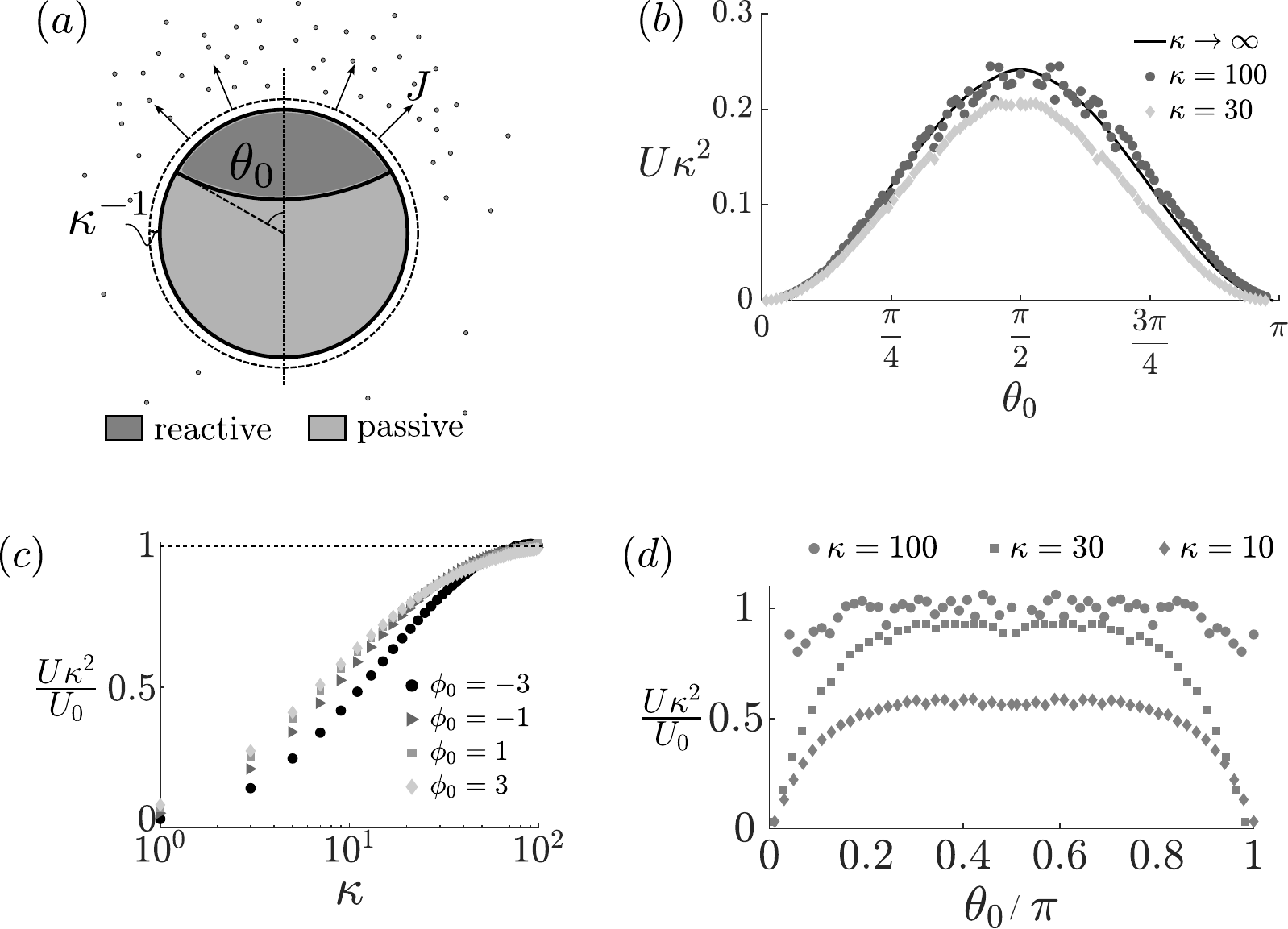}
    \caption{(a) Self-phoretic Janus particle where propulsion is controlled by the size of the spherical cap $\theta_0$, the reactive flux $J$ and the interaction length scale $\kappa^{-1}$. (b) $U$ vs $\theta_0$ for different $\kappa$ values demonstrates a maximum velocity for $\theta_0=\frac{\pi}{2}$ irrespective of $\kappa$. (c) $\frac{U\kappa^2}{U_0}$ versus $\kappa$ asymptotically approaches the thin limit calculations as $\kappa \to 100$. However, considerable dampening is observed even for $\kappa = O(10)$. The values are reported for $\theta_0=\frac{\pi}{2}$. (d) The dampening of $\frac{U\kappa^2}{U_0}$ with $\kappa$ is observed for all $\theta_0$. The value $U_0$ is the asymptotic limit of $U \kappa^2$ from the thin interaction layer calculations.  } 
    \label{Fig: hem-cap-res}
\end{figure}

 In this section, we use the formula obtained in Eqs. \eqref{Eq: unified_trans_exp_for_spheres}-\eqref{Eq: unified_rot_exp_for_spheres} to study the translation of Janus-like particles with a spherical cap, see Fig.\ref{Fig: hem-cap-res}a. The key novelty of our analysis is that Eqs. \eqref{Eq: unified_trans_exp_for_spheres}-\eqref{Eq: unified_rot_exp_for_spheres} do not impose the restriction on interaction lengthscale. As we show later, if the interaction length is comparable to particle size, the particle velocity is significantly impacted. 
 
 We define catalytic surface activity through a non-dimensional outward surface flux of strength, $J$ (scaled by a characteristic flux $D c_0 / a$, where $D$ is the diffusivity of the solute, $c_0$ is a reference concentration of the solute and $a$ is the particle radius). The non-dimensional interaction length is characterized by $\kappa^{-1}$ (scaled by $a$). Lastly, the size of the catalytic cap is controlled by the polar angle $\theta_0$, such that $\theta_0=0$ indicates no catalytic cap on the particle and $\theta_0=\frac{\pi}{2}$ represents a hemispherical cap. 
 
We assume a Helmholtz-like equation governs the interaction potential ($\phi$, scaled by $k_B T$) with a constant surface potential ($\phi_0$) and a far-field decay. We take this opportunity to highlight the choice of Helmholtz-like potential. While the potential is not representative of different surface interactions possible, it provides a convenient choice to explore the impact of $\kappa$ and thus has been chosen for this analysis. We note that our analysis can be easily extended to other interaction potentials provided that the integrals in Eqs. \eqref{Eq: unified_mobility_expression} are convergent. For a detailed analysis of phoretic motion due to a general particle-solute interaction the reader is directed to \cite{brady2021phoretic}.

To resolve particle translation for a given surface activity and interaction, we write

\begin{equation}
    \nabla^2 \phi = \kappa^2 \phi, \label{Eq: Poissons Eq potential}
\end{equation}
\begin{equation}
    \phi = \phi_0,\;\;\;r=1, \label{Eq: surface potential}
\end{equation}
\begin{equation}
     \phi \to 0,\;\;\;r \to \infty. \label{Eq: potential decay far-field}
 \end{equation}

\noindent Solute transport is governed by diffusion and phoretic interactions with the following boundary conditions

\begin{equation}
        \bm{\nabla} \cdot \left(\bm{\nabla} c + c \bm{\nabla} \phi \right) = 0, \label{Eq: solute transport}
\end{equation}
\begin{equation}
        -\mathbf{n}\cdot\left(\bm{\nabla} c + c\bm{\nabla}\phi\right) = J,\;\;\;r=1, \label{Eq: solute surface boundary condition}
\end{equation}
\begin{equation}
        c\to0,\;\;\;r\to\infty. \label{Eq: solute far-field condition}
\end{equation}

\noindent Our model problem is illustrated in Fig. \ref{Fig: hem-cap-res}(a). We solve the coupled equations \eqref{Eq: Poissons Eq potential}-\eqref{Eq: solute far-field condition} numerically. A scaled model geometry was constructed with the particle radius given to be $a=1$ and an outer radius of $r|_\infty=20$, representing the far field. The interaction potential ($\phi$) and solute concentration ($c$) were defined by Eq. \eqref{Eq: Poissons Eq potential}-\eqref{Eq: solute far-field condition}. To obtain the translation velocity through Eq. \eqref{Eq: unified_trans_exp_for_spheres}, we define the body force $\mathbf{b} = -c\bm{\nabla}\phi-\bm{\nabla} c$. Note that the arbitrary body force has both a phoretic and an osmotic contribution. Details of the computational method are provided in Appendix B.

We first analyze the effects of the spherical cap size ($\theta_0$) and interaction length scale ($\kappa^{-1}$) for a fixed surface flux ($J=1$) and surface potential ($\phi_0=-1$); see Fig.~\ref{Fig: hem-cap-res}. First, we note that the particle moves in the direction of the catalytic cap, since when $\phi_0=-1$ the particle is attracted towards regions of higher solute concentration. From Fig. \ref{Fig: hem-cap-res}(b), we observe that the propulsion velocity is maximum for $\theta_0=\frac{\pi}{2}$. This is in agreement with prior observations in the literature, \cite{golestanian2007designing, michelin2014phoretic,popescu2018effective} and others. Additionally, the dependence of translation velocity on the cap size is symmetric about $\theta_0=\frac{\pi}{2}$. As described in \cite{michelin2014phoretic}, when $\theta_0=\frac{\pi}{2}$, for small P\'{e}clet numbers, the sharpest concentration gradients are located near the equator, consequently leading to a larger slip velocity over an extended inert surface. In contrast, for smaller or larger catalytic caps, the aforementioned solute front is closer to the pole and thus involves a smaller share of the particle surface in generating slip velocities and therefore a smaller swimming speed. 

\noindent\textbf{Thin interaction layer:} We also study the dependence of propulsion velocity on the interaction length scale ($\kappa^{-1}$); see Fig.~\ref{Fig: hem-cap-res}(c). To compare our calculations with the thin interaction layer limit, we perform analogous calculations following the approach of  \cite{anderson1989colloid} and \cite{derjaguin1947kinetic}. We consider diffusive transport of solute $\nabla^2 c = 0$ through the fluid volume. The surface flux condition is given by $-\mathbf{n}\cdot \bm{\nabla}c=J$ with a far-field decay condition. The phoretic slip at the particle surface is defined as,
\begin{equation}
    \mathbf{u}_\mathrm{slip} = - \bm{\nabla}_s c \displaystyle \int_1^\infty (r-1)\left[\exp\left(-\phi\right)-1\right] \ dr, \label{Eq: slip velocity calculation}
\end{equation}
\noindent where the non-dimensional $\mathbf{u}_\mathrm{slip}$ is scaled with $k T c_0 a / \mu$. After that, we refer to \cite{stone1996propulsion}, also derived in Eq. \eqref{Eq: stone and samuel translation}, to obtain the translation velocity $\mathbf{U}^{\rm thin}$ to be,
\begin{equation}
	\mathbf{U}_\mathrm{thin} = -\frac{1}{4\pi} \displaystyle \int_S \mathbf{u}_\mathrm{slip} {\rm d}S. \label{Eq: stone and samuel result}
\end{equation}
It is well-known in diffusiophoretic literature \citep{golestanian2019phoretic} that for $\kappa \gg 1$, $|\mathbf{U}_\mathrm{thin}| \propto 1/\kappa_\mathrm{thin}^2$. Consequently, we introduce a scaled velocity expression, $\mathbf{U}_0 = \mathbf{U}_\mathrm{thin} \kappa^2_\mathrm{thin}$ which becomes constant as $\kappa \to \infty$. As shown in Fig. \ref{Fig: hem-cap-res}(c), we observe that the velocity ratio $U\kappa^2/U_0 = |\mathbf{U}|\kappa^2/|\mathbf{U}_0|$ significantly decreases when the interaction limit becomes comparable to particle size, i.e., $\kappa \rightarrow 1$. We find that even for $\kappa = O(10)$,  the velocity is reduced by almost a factor of 2.  We only reach the thin interaction limit for $\kappa = O(10^2)$. This observation is consistent with passive diffusiophoretic literature where even moderately thin double layers can significantly reduce the diffusiophoretic velocity \citep{prieve1984motion}; see Fig. 3(b). Our analysis highlights that even for autophoretic swimmers, this effect can be observed, and using a thin interaction limit could overestimate the velocity for moderately thin interaction thickness such as $\kappa \rightarrow 50$. Over the past decade, there has been an increasing interest in nanoparticles or very dilute systems where relative interaction lengthscales are large \citep{wu2021mechanisms,shi2023nanomotor,shin2016size, gupta2020diffusiophoresis,leunissen2007electrostatics}. Thus these results could be crucial for future experimental studies. To ensure that our trends of reductions in velocity are consistent for other conditions, in Fig. \ref{Fig: hem-cap-res}(d), we observe that reduction in velocity for smaller $\kappa$ values is consistent for all $\theta_0$ values. 

\section{Autophoretic swimmers with external solute gradients}
\begin{figure}
    \centering
    \includegraphics[width=\textwidth]{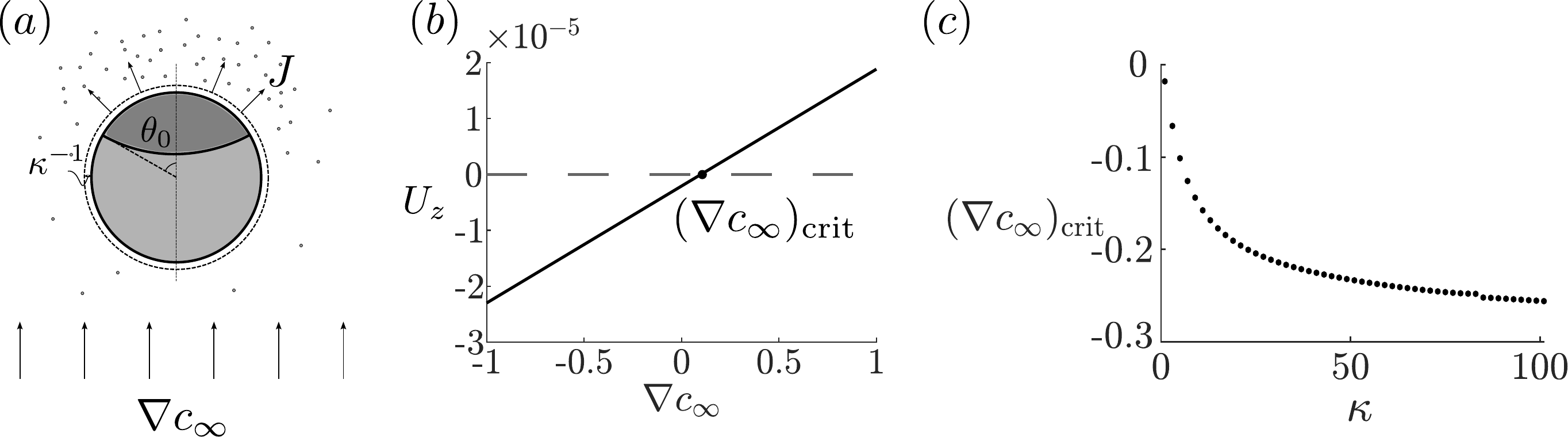}
    \caption{(a) A self-phoretic particle also driven by external $\nabla c_\infty$. (b) $U_z$ vs. $\nabla c_\infty$ shows that both the particle speed and direction depend on the competing effects of external and self-propulsion modes. We define $(\nabla c_\infty)_\mathrm{crit}$ as the value when the particle motion was arrested despite the presence of concentration gradients due to surface activities. (c) $(\nabla c_\infty)_\mathrm{crit}$ vs $\kappa$. We observe an increase in the magnitude of the external flux needed to arrest motion as we approach thin interaction length limits.} 
    \label{Fig: hem-cap-res-2}
\end{figure}

 To further demonstrate the applicability of Eq. \eqref{Eq: unified_trans_exp_for_spheres} in scenarios where it is difficult to use a slip velocity approach, we modify our model problem to include an external solute flux in the fluid bulk. 
 The model geometry is preserved as shown in Fig. \ref{Fig: hem-cap-res-2}a. The concentration in the far field is $\bm{\nabla} c_\infty = J_\mathrm{ext} \mathbf{e}_z$, where $\mathbf{e}_z$ is the $z$-coordinate basis vector in the universal Cartesian frame of reference and $J_\mathrm{ext}$ is a free parameter used to control the direction and strength of this external field. The magnitude of the external gradient is thus $\nabla c_\infty = J_\mathrm{ext}$. We note that all additional parameters have been appropriately non-dimensionalized as per the discussion in section \ref{sec: autophoretic motion of microswimmers}. We analyze the variations of the local surface flux ($J$), external solute flux ($J_\mathrm{ext}$), and interaction length scale $\kappa$. 
 For a given surface flux $J=1$, it is observed that the translation velocity is linear with the imposed solute flux, $\nabla c_\infty$. The particle moves with a velocity $\mathbf{U} = U_z \mathbf{e}_z$. The direction of the propulsion is governed by the relative magnitudes of different contributions of the body force terms. In the absence of an external flux, the body force contributions due to the phoretic activity cause the particle to move in the positive $z$-direction. In our analysis, we have ignored the excess osmotic contribution ($\nabla c$) from the external solute gradient and only considered the phoretic contribution. This is done to ignore the motion of the particle only due to $\nabla c_\infty$ when $\phi=0$. For $J_\mathrm{ext}>0$, one can observe that force terms arising from the gradients in active and passive concentration fields are in the same direction, hence $U_z > 0$ as in Fig. \ref{Fig: hem-cap-res-2}(b). Alternatively when $J_\mathrm{ext} < 0$, the phoretic force terms arising out of the passive solute concentration field compete with forcing arising due to activity. This leads to a direction reversal below a threshold concentration gradient, $(\nabla c_\infty)_\mathrm{crit}$, where $U_z < 0$. 
 $(\nabla c_\infty)_\mathrm{crit}$ is the external solute gradient necessary to arrest particle motion due to activity. The linearity of the results in Fig. \ref{Fig: hem-cap-res-2}b is due to the computations being performed in the weak field limit. 
 To inspect the effects of interaction potential length, we obtained the  $(\nabla c_\infty)_\mathrm{crit}$ values at $J=1$ while varying over $\kappa$. A qualitative agreement is observed between the results in Fig. \ref{Fig: hem-cap-res-2}(c) and the $\kappa$-dependence of velocity ($U$). Further, we see that a stronger external flux is necessary to arrest motion as we approach the thin interaction length limit. This effect arises because, for $\kappa \to \infty$, the catalytically ejected solute decays more sharply, $\nabla c$ increases locally near the particle surface. This necessitates the need for a larger $(\nabla c_\infty)_\mathrm{crit}$ to counteract activity-induced gradients to arrest propulsion. 

\label{sec:swimming}
\section{Conclusion} \label{sec: conclusion}
The main result of our work is that \eqref{Eq: unified_trans_exp_for_spheres}-\eqref{Eq: unified_rot_exp_for_spheres}, reported in prior literature for phoretic systems~\citep{brady2021phoretic} and different physical setups \citep{brenner1963resistance, hinch1991, leal2007advanced,datt2015squirming, elfring2016effect,datt2017active}, can retrieve the mobility expressions for both electrophoretic and electrolytic diffusiophoretic motion of colloidal spherical particles at arbitrary interaction lengths at the low potential limit. Additionally, the asymptotic limit of equations \eqref{Eq: unified_trans_exp_for_spheres}-\eqref{Eq: unified_rot_exp_for_spheres} for thin interaction length recovers net translation and rotation of the particle in terms of a phoretic slip velocity as in \cite{stone1996propulsion}. Finally, we employ these mobility relationships for self-propulsion of spherical microswimmers where we observe a peak in translational velocity as we approach a hemispherical catalytic coverage with a velocity reduction for lower or higher coverages. At moderate interaction lengths, $\kappa = O(10)$, a dampening in the translation velocity is observed. However, for higher $\kappa$ values our calculations retrieve the thin interaction length limit results arising from equivalent slip velocity calculations. Further, we add an external passive solute concentration gradient to our problem to understand the competing effects of surface-generated and externally imposed solute concentration gradients. We find configurations where the external flux arrest motion induced by surface activity ($\nabla c_\infty)_\mathrm{crit}$ and regions where the propulsion induced by the passive external solute gradient aids with or competes against the propulsion induced due to surface activity. A non-linear decay of $(\nabla c_\infty)_\mathrm{crit}$ with $\kappa$ is observed from our analysis which is qualitatively similar to the $\kappa$-dependence of the translation velocity, $|\mathbf{U}|$.
\par{} Beyond the result described in this paper, our work can be utilized to predict $\mathbf{U}$ and $\mathbf{\Omega}$ for an arbitrarily shaped particle. For such a calculation, one would need the appropriate expressions of the mobility ($\mathbf{M}$) and the disturbance tensors ($\mathbf{D}_T$, $\mathbf{D}_R$), which might be possible to obtain analytically or numerically.  While this manuscript focuses on the mobility of a single particle, a similar analysis could be extended to multiple particles and particles under confinement.   
\par{} The derived expression will be particularly useful for multiphysics propulsion. For instance, it might be possible to induce propulsion of particles using multiple modes such as a combination of electric fields and other fields, such as electrodiffusiophoresis \citep{wang2022visualization, jarvey2023asymmetric}. Another possibility is the inclusion of chemical kinetics at the particle surface \citep{davis2022self} which will modify the solute problem and thus consequently change $\mathbf{b}$.

\section*{Acknowledgement}
The authors acknowledge H.A. Stone, A.S. Khair, F. Henrique, G. Roure, and the three anonymous referees for their input in the preparation of this manuscript. Ankur G. thanks the National Science Foundation (CBET - 2238412) CAREER award for financial support. Arkava G. acknowledges the Teets Family Endowment in Nano-Technology Graduate Fellowship and the Mukhopadhyay Graduate Research Fellowship. Souradeep R. acknowledges the Mukhopadhyay Graduate Research Fellowship. The authors acknowledge the donors of the American Chemical Society Petroleum Research Fund.

\section*{Declaration of Interests}
The authors report no conflict of interest.

\appendix
\label{appendix:a}
\section{Numerical resolution of the radial components at $O(\epsilon\zeta)$ and $O(\epsilon \zeta^2)$}

\makeatletter 
\renewcommand{\theequation}{A\@arabic\c@equation}
\makeatother
\setcounter{equation}{0}

$\bm{\mathrm{O}(\epsilon \zeta)}$:The radial components of Eqs. \eqref{Eq: diff_ord_11_chrg_bal}-\eqref{Eq: diff_ord_11_surf_potential} are,
\begin{equation}
    \begin{aligned}
        \frac{1}{r^2}\frac{d}{dr}\left(r^2\frac{d f_{\rho_{11}}}{dr}\right) - \left(\frac{2}{r^2} + \kappa^2\right)  f_{\rho_{11}} + \frac{2}{r^2} \frac{d}{dr}\left(r^2 f_{s_{10}}(r) \frac{d \phi_{01}(r)}{d r}\right) =0, \label{Eq: diff_ord_11_rho_radial_component}
    \end{aligned}
\end{equation}
\begin{equation}
    \frac{1}{r^2}\frac{d}{dr}\left(r^2 \frac{df_{s_{11}}}{dr}\right)-\frac{2f_{s_{11}}}{r^2}+\frac{1}{r^2}\frac{d}{dr}\left(r^2\rho_{01}(r)\frac{ df_{\phi_{10}}(r)}{dr}\right) - \frac{2 \rho_{01}(r) f_{\phi_{10}}(r)}{r^2}=0, \label{Eq: diff_ord_11_salt_radial_component}
\end{equation}
\begin{equation}
    \frac{1}{r^2}\frac{d}{dr}\left(r^2\frac{df_{\phi_{11}}}{dr}\right)-\frac{2f_{\phi_{11}}}{r^2} = -\frac{\kappa^2}{2}f_{\rho_{11}}, \label{Eq: diff_ord_11_poisson_eq_radial_component}
\end{equation}
\noindent where $f_{\phi_{10}}(r)$ and $f_{s_{10}}(r)$ are the radial components of the $O(\epsilon)$ solutions. The appropriate boundary conditions at the particle surface are,
\begin{equation}
    \frac{df_{\rho_{11}}}{dr} + f_{s_{10}}(r) \frac{d \phi_{01}(r)}{d r}=0, \label{Eq: diff_ord_11_chrg_flx_radial_component}
\end{equation}
\begin{equation}
    \frac{df_{s_{11}}}{dr} + \rho_{01}(r)\frac{d f_{\phi_{10}}(r)}{d r}=0, \label{Eq: diff_ord_11_surf_salt_flx_radial_component}
\end{equation}
\begin{equation}
    \frac{\mathrm{d}f_{\phi_{11}}}{\mathrm{d}r} = 0 \label{Eq: diff_ord_11_potential_flx_radial_component}
\end{equation}
In the far field, the radial flux of $f_{\phi_{11}}$, $f_{\rho_{11}}$, and $f_{s_{11}}$ all go to zero.

$\bm{\mathrm{O}(\epsilon \zeta^2)}$: The radial dependence of $s_{12}(r, \theta)$ is captured by,
\begin{equation}
    \frac{1}{r^2}\frac{d}{dr}\left(r^2\frac{d f_{s_{12}}}{dr}\right)-\frac{2f_{s_{12}}}{r^2}+\frac{1}{r^2}\frac{d}{dr}\left(r^2\left\{\rho_{01}(r) \frac{d f_{\phi_{11}}(r)}{d r} + f_{\rho_{11}}(r) \frac{d f_{\phi_{01}}(r)}{d r}\right\}\right) - \frac{2 f_{\phi_{11}}\rho_{01}}{r^2} = 0, \label{Eq: diff_ord_12_radial_salt_bal}
\end{equation}
with the boundary condition at the particle surface being
\begin{equation}
    \frac{df_{s_{12}}}{dr}+\rho_{01} \frac{d f_{\phi_{11}}}{d r} + f_{\rho_{11}} \frac{d \phi_{01}}{d r}=0, \label{Eq: diff_ord_12_surf_radial_salt_flx}
\end{equation}
and $f_{s_{12}}=0$ in the far-field. We solve Eqs. \eqref{Eq: diff_ord_11_rho_radial_component}-\eqref{Eq: diff_ord_12_radial_salt_bal} using the \emph{bvp4c()} function in MATLAB for $\kappa \in \left[1, 1000\right]$. For each $\kappa$ value we solve for $\rho$, $\phi$, and $s$ at each order with a one-dimensional mesh, $r = [1, 1+100/\kappa]$ with a thousand elements.  A default relative tolerance of $10^{-3}$ is used. To match with Eq. \eqref{Eq: diff_vel_keh} we multiply the terms proportional to $\zeta$ in Eq. \eqref{Eq: diffusiophoretic mobility} with a factor of $1/2$ and the terms proportional to $\zeta^2$ with a factor of $4$. This is solely due to how the coefficients $\Theta_1(\kappa)$ and $\Theta_2(\kappa)$ are defined in Eq. \eqref{Eq: diff_vel_keh}.

\label{appendix:b}
\section{Numerical solution to the autophoretic motion of microswimmers in sections 4 and 5}

\makeatletter 
\renewcommand{\theequation}{B\@arabic\c@equation}
\makeatother
\setcounter{equation}{0}
\noindent\textbf{Section 4:} The numerical solutions to Eqs. \eqref{Eq: Poissons Eq potential} to \eqref{Eq: solute far-field condition} were obtained from the finite element method software COMSOL. We use a non-dimensional spherical computational domain of size $4/3\pi\times20^3$ where the particle is located at the origin and possesses a radius of unity. The domain is discretized into approximately 1,162,711 elements. Around the particle, we mesh a boundary region with 12 layers and a stretching factor of 1.1, consisting of approximately 20,000 triangular elements. The simulations are performed in the reference frame of the swimmer. Upon solving for the solute concentration $c$, and $\phi$ in the domain we define $\mathbf{b} = -c\bm{\nabla}\phi-\bm{\nabla}c$ as the body force. Subsequently, Eq. \eqref{Eq: unified_trans_exp_for_spheres} is numerically integrated over the domain to find the translation velocity $\mathbf{U}$. All results are normalized with $\kappa^2$ as illustrated in Figs. \ref{Fig: hem-cap-res} and \ref{Fig: hem-cap-res-2}. \\
\noindent The translation velocity at the thin double layer limit is obtained by a similar procedure by solving for $c$ and using Eq. \eqref{Eq: slip velocity calculation} to obtain $\mathbf{u}_{\rm slip}$. For the second simulation setup, to calculate the velocity at the thin limit an extremely fine mesh is required to discretize the domain, with approximately 1,154,140 elements. The lumped phoretic mobility for a given surface potential $\phi$ is obtained by numerically integrating along the radial direction in MATLAB from $1$ to approximately $10^4$ along the radial direction. The translation velocity $\mathbf{U}_{\rm thin}$ is obtained by solving Eq. \eqref{Eq: stone and samuel result} in COMSOL.

\noindent\textbf{Section 5:} To solve the results where the swimmer was subjected to an external concentration gradient, we modify the far-field boundary condition in the above numerical simulation to be $c_\infty = c_0 + J_{\rm ext}z$. A Dirichlet boundary condition was used instead of a Neumann boundary condition to avoid under-specifying our model. The Dirichlet condition will satisfactorily approximate $\bm{\nabla}c_\infty = J_{\rm ext}\mathbf{e}_z$ in the vicinity of the particle for a large enough computational domain.
\bibliographystyle{jfm}
\bibliography{jfm}

\end{document}